\documentclass[aps,prb,twocolumn]{revtex4-2}
\usepackage{amsmath}
\usepackage{amssymb}
\usepackage{mathtools}
\usepackage{graphicx}
\usepackage{tabularx}
\usepackage{bm}
\usepackage{amsthm}
\usepackage[toc,page]{appendix}
\usepackage{caption}
\usepackage{subcaption}
\usepackage{ragged2e}
\usepackage{multirow}
\usepackage{enumitem}
\usepackage{hyperref}
\usepackage{xcolor}
\usepackage[normalem]{ulem}

\newcommand{\curl}[1]{\nabla\times #1}

\makeatletter
\renewcommand{\maketag@@@}[1]{\hbox{\m@th\normalsize\normalfont#1}}%
\makeatother

\begin{document}

\title{A Time-Dependent Ginzburg-Landau Framework for Sample-Specific Simulation of Superconductors for SRF Applications}

\author{Aiden V. Harbick}
\email{aharbick@student.byu.edu}
\author{Mark K. Transtrum}
\email{mktranstrum@byu.edu}
\affiliation{Department of Physics and Astronomy, Brigham Young University, Provo, Utah 84602, USA}

\date{June 19, 2025}

\begin{abstract}
  Modern superconducting radio frequency (SRF) applications demand precise control over material properties across multiple length scales—from microscopic composition, to mesoscopic defect structures, to macroscopic cavity geometry. We present a time-dependent Ginzburg-Landau (TDGL) framework that incorporates spatially varying parameters derived from experimental measurements and ab initio calculations, enabling realistic, sample-specific simulations. As a demonstration, we model Sn-deficient islands in Nb\textsubscript{3}Sn and calculate the field at which vortex nucleation first occurs for various defect configurations. These thresholds serve as a predictive tool for identifying defects likely to degrade SRF cavity performance. We then simulate the resulting dissipation and show how aggregate contributions from multiple small defects can reproduce trends consistent with high-field $Q$-slope behavior observed experimentally. Our results offer a pathway for connecting microscopic defect properties to macroscopic SRF performance using a computationally efficient mesoscopic model.

\end{abstract}

\maketitle

\section{Introduction}
Superconducting Radio-Frequency (SRF) cavities are a crucial component of particle accelerators, as they utilize AC electromagnetic fields to accelerate beams of charged particles. Nb has been the industry standard for SRF applications for decades due to its high critical temperature ($\sim$9K) relative to the other elemental superconductors. Within the past decade, the need for SRF cavities with capabilities beyond the limits of Nb cavity performance has led to the study of a variety of alternative SRF materials. Among these materials, Nb\textsubscript{3}Sn has emerged as a promising candidate; Nb\textsubscript{3}Sn boasts both a higher critical temperature ($\sim$18K) and higher critical fields \cite{Posen_Nb3Sn_2014,Posen_Nb3Sn_2017}. One particular advantage of SRF cavities (as compared to traditional normal conducting RF cavities) is their high quality factors (Q) \cite{gurevich_RF_Superconductivity_Review}. A major benefit of Nb\textsubscript{3}Sn SRF cavities compared to their Nb counterparts is that they can maintain similar Qs (on the order of $10^{10}$) at higher temperatures (4.2K vs. 2K) \cite{Posen_Nb3Sn_2017}, significantly reducing cryogenic costs. NbZr is another promising alternative SRF material which has seen recent attention \cite{Sun_NbZr,Sitaraman_NbZr}, with most existing NbZr samples exhibiting critical temperatures between 10 and 13K (the theoretical maximum is 17.7K), but this material has not yet been tested at cavity scale. The simulations in this paper focus on Nb\textsubscript{3}Sn, but the methods we will present can be generalized to any material of interest. 

The oscillating electric fields used for acceleration in SRF cavities induce magnetic fields parallel to the cavity surface, so for large accelerating gradients, the critical magnetic fields of the cavity material are the fundamental limits on cavity performance. Type-II superconductors such as Nb and Nb\textsubscript{3}Sn have two critical fields, $H_{c1}$ and $H_{c2}$. For fields below $H_{c1}$, the Meissner state is stable and magnetic flux is expelled from the cavity. Between $H_{c1}$ and $H_{c2}$, there is a mixed state in which superconducting vortices trap lines of magnetic flux, forming normal cores inside the otherwise superconducting state. Under an AC field, as is the case for SRF cavities, the vortices quickly move in and out of the cavity over the course of an AC cycle. This vortex motion leads to large amounts of dissipation \cite{Bardeen_Stephen}. For fields above $H_{c2}$, the mixed state becomes unstable and in this state SRF cavities will quench (i.e. go normal conducting). It is important to note that this is not the only mechanism for SRF cavity quench, the dissipation caused by moving vortices in the mixed state can cause heating in the cavity, which can also lead to quenching through a change in cavity temperature. As such, for SRF applications, it is important that the cavity remain within the Meissner state during operation. 

While the Meissner state is no longer thermodynamically stable above $H_{c1}$, it can remain metastable up until the so-called superheating field, $H_{sh}$ \cite{Liarte_Hsh}. It is well known that many high-power SRF cavities operate in this metastable Meissner state \cite{Posen_RF_Limits}. As such, $H_{sh}$ is the theoretical limiting field for operation of SRF cavities, since the dissipative vortices which are detrimental to SRF performance become unavoidable for fields above $H_{sh}$. $H_{sh}$ has been studied for decades by condensed matter theorists. These studies have most commonly been within a Ginzburg-Landau (GL) framework \cite{Kramer_Hsh,Dolgert_Hsh,Transtrum_Hsh,Hsh_anisotropy}, but the superheating field has also been studied extensively utilizing the Eilenberger equations \cite{Catelani_Eilenberger_Hsh, Lin_Gurevich_Eilenberger_Hsh, Kubo_Quasiclasical_Hsh}. 

$H_{sh}$ provides the maximum possible field (and therefore the maximum accelerating gradient) for SRF cavity operation, but local features of a material such as impurities or surface geometries can act as nucleation sites for vortices. This means that in practice, realistic material samples will be limited by what we will call the vortex penetration field, $H_{vort}$, which is the lowest field at which the material nucleates vortices. This quantity can vary greatly between different samples and depends on a large variety of different effects, so estimation of $H_{vort}$ for realistic sample materials remains a rich area of research. For the case of Nb\textsubscript{3}Sn in particular, theoretical $H_{sh}$ calculations suggest that Nb\textsubscript{3}Sn cavities could reach accelerating gradients as high as around 100 MV/m, yet the highest accelerating field achieved so far by a Nb\textsubscript{3}Sn SRF cavity is around 24 MV/m \cite{Posen_Best_Cavity}, with most other cavities reaching their quench field well below this. Additionally, many existing cavities exhibit a phenomenon in which Q significantly degrades as the cavity approaches its quench field, a phenomenon dubbed ``Q-slope" \cite{Q_slope}, also sometimes called high field Q-slope (HFQS) when it occurs primarily at higher fields near the quench field. These performance degradations are the result of material defects introduced during the Nb\textsubscript{3}Sn growth process, so understanding how different defects seen within samples affect things like $H_{vort}$ or dissipation more generally is critical to developing better growth techniques. 

The need to accurately model specific features such as material impurities within superconducting materials motivates us to develop a framework which will allow us to directly model the spatial variations of superconducting properties due to the material compositions observed in realistic sample materials. To do this, we use time-dependent Ginzburg-Landau (TDGL) theory. TDGL has already proven itself to be a powerful tool for mesoscopic-scale simulations relevant to SRF applications \cite{Pack_Vortex_Nucleation,Carlson_GBs,Oripov_Theory,Oripov_Applied}. Besides SRF simulations, TDGL has broad application such as in single photon detectors \cite{Single_Photon_Detector_TDGL,Single_Photon_Detector_GTDGL}, superconducting quantum interference devices (SQUIDs) \cite{SQUIDs,BishopVanHorn_pyTDGL}, weak links \cite{TDGL_Weak_Links,TDGL_Weak_Links_2,Weak_Link_Chemical_Potential,Weak_links_magnetoresistance,Phase_Slips_Weak_Links}, or superconducting nanowires \cite{Nanowires_3dFE,Nanowires_theory,Nanowires_PhaseSlips}. The methods we present in this paper are investigated with full 3D simulations. While prior work has studied vortex dynamics in 3D \cite{petukhov1973rate, burlachkov1993magnetic,feinberg1994vortex,guo2015extracting,sadovskyy2016simulation,guo2017situ,zhakina2024vortex}, the majority of TDGL research, especially in the context of SRF cavities, has been limited to 2D.

Whenever working with TDGL, it is important to acknowledge its limitations. TDGL is a mesoscopic-scale model which abstracts the microscopic details of superconductivity into quantities which can be used to describe things like vortex dynamics. Much of this abstraction is the direct result of restricting the quantitative validity to gapless superconductivity at temperatures near $T_c$. This means that outside of this fairly restrictive regime, the quantitative predictions of TDGL are not accurate in general. Despite this, there are three main reasons that TDGL is still has considerable value in a variety of studies, such as those referenced above. Firstly, it is well known that the solutions to the Ginzburg-Landau equations have a much wider range of quantitative applicability when under the dirty limit, and so properties of the TDGL equations which can be derived from steady-state dirty limit solutions, such as $H_{sh}$, are still valuable quantitative outputs of the theory. Secondly, theories of superconductivity with larger ranges of quantitative accuracy (such as the Eilenberger equations \cite{Catelani_Eilenberger_Hsh, Lin_Gurevich_Eilenberger_Hsh, Kubo_Quasiclasical_Hsh} or other quasi-classical approaches) scale extremely poorly in terms of computational complexity when it comes to numerical simulations, whereas TDGL is at least feasible for larger mesoscopic-scale simulations. And finally, TDGL offers qualitative and semi-quantitative predictions that provide useful insight into phenomena which may be difficult or impossible to measure experimentally. For example, it can be used to compare how variations in the size and depth of stoichiometric defects relative to the superconductor’s surface affect SRF-relevant metrics such as vortex nucleation and energy dissipation---enabling prioritization of which defect characteristics are most critical to address when perfect control is not possible. In what follows, we apply this approach to a specific class of such defects relevant to Nb\textsubscript{3}Sn-coated SRF cavities.

To model sample-specific materials and investigate mechanisms behind Q-slope and other quenching phenomena in Nb\textsubscript{3}Sn SRF cavities, we draw on both experimental and theoretical studies that characterize the microscopic properties of Nb\textsubscript{3}Sn. There has been a large body of work experimentally characterizing SRF grade vapor-diffused Nb\textsubscript{3}Sn samples. The primary suspect for SRF performance degradation comes from defects or other imperfections in the Nb\textsubscript{3}Sn surface significantly lowering the barrier to flux penetration \cite{Posen_Valles_Liepe_field_limits}. In particular, defects which have been studied are abnormally thin or patchy grains \cite{Trenikhina_Patchy_Grains_Sn_islands,Lee_Patchy_Grains_Sn_islands,Pudasaini_Patchy_Grains}, Sn-segregated grain boundaries \cite{Lee_Sn_Segregated_GBs,Oh_Diffusion_GBs,Carlson_GBs}, and Sn-deficient regions \cite{Trenikhina_Patchy_Grains_Sn_islands,Lee_Patchy_Grains_Sn_islands,Becker_Sn_islands,Viklund_Sn_Islands}. In addition to experimental characterizations, there have also been a variety of \textit{ab initio} calculations for Nb\textsubscript{3}Sn using density functional theory. In addition to calculations of general properties of Nb\textsubscript{3}Sn \cite{Freericks_DFT_Nb3Sn}, such as the electron and phonon density of states and Eliashberg spectral function, these quantities have also been estimated with respect to varying intrinsic strain \cite{MARKIEWICZ_DFT_Nb3Sn_Strain} as well as normal resistivity \cite{Mentink_DFT_Normal_Resistivity}. Variations in the superconducting $T_c$ as well as electron density of states have also been calculated with respect to varying tin concentration \cite{Kelley_GBs,Sitaraman_Nb3Sn}, which applies to both Sn-segregation at grain boundaries and Sn-deficient regions. These Sn-deficient regions, which we will call Sn-deficient islands, are the primary material defect we will study in order to validate our methods.

Experimental characterizations can give data about the material compositions and physical structure of superconducting materials, and \textit{ab initio} calculations provide detailed descriptions of the electronic/phononic structure and the resulting superconducting properties, both of these a microscopic scale. TDGL plays the role of modeling mesoscopic scale phenomena (such as vortex dynamics) which are difficult or even impossible to measure directly via experiment, and are too large to easily model with microscopic scale theories such as DFT. There have been a number of studies which have used TDGL to model material inhomogenieties \cite{Koshelev_Pinning,Sadovskyy_Pinning,Sorensen_Pinning,Chapman_Pinning,Pack_Vortex_Nucleation,Carlson_GBs,AlLuhaibi_Vortex_Hopping}, but these studies did not use the explicit dependencies of the TDGL parameters on microscopic material properties to inform their choice of parameters. The limitation to the approach used in these references is that it requires either looking through a large portion of the TDGL parameter space in order to find values which lead to expected predictions, or more commonly, picking somewhat arbitrary values, which limits confidence in the results.  

In this paper, we outline a new framework in TDGL theory which allows us to directly calculate the values of the TDGL parameters based on local properties of the superconductor. This framework enables modeling of realistic features of superconductor samples and supports estimation of critical fields and energy dissipation under dynamic electromagnetic conditions. Under our framework, TDGL serves as a bridge between experimental material characterizations and \textit{ab initio} calculations of material-specific parameters, allowing us to further connect these microscopic characterizations with macroscopic SRF performance metrics in a sample-specific way. Because SRF cavity development is inherently multidisciplinary---bringing together accelerator physicists, materials scientists, and condensed matter theorists---a framework that integrates insights across these domains is particularly valuable. The method presented here enables such integration, offering a pathway to sample-specific predictions grounded in both microscopic characterization and macroscopic application.

This paper is organized as follows. In Section \ref{Methods}, we present the TDGL equations and show how to calculate TDGL parameters from material properties. We then describe how spatial variations in these parameters are estimated by combining results from DFT calculations with experimental material characterizations. The section concludes with a discussion of how dissipation can be computed from TDGL solutions and how SRF cavity quality factors can be derived from this dissipation. In Section \ref{Results}, we apply our framework to estimate the critical fields associated with Sn-deficient islands of varying size, position, and stoichiometric composition. We then examine dissipation in the context of Nb SRF cavities and analyze a representative simulation in detail to probe the limits of TDGL. Finally, we demonstrate how combining critical field and dissipation results enables us to estimate high-field Q-slope behavior for Nb\textsubscript{3}Sn SRF cavities. Section \ref{Conclusions} concludes the paper with a discussion of the justification for using TDGL despite its known limitations and the implications of our findings for future SRF cavity research.

\section{Methods}\label{Methods}

\subsection{The Time-Dependent Ginzburg-Landau Equations}\label{TDGL}

Ginzburg-Landau (GL) theory is one of the oldest theories of superconductivity, and it remains relevant today owing to its relative simplicity and direct physical insights into the electrodynamic response of superconductors under static applied fields and currents \cite{GL_ref}. The \textit{time-dependent} Ginzburg-Landau (TDGL) equations were originally proposed by Schmid \cite{Schmid_TDGL} in 1966 and Gor'kov and Eliashberg \cite{GorKov_TDGL} derived them rigorously from BCS theory later in 1968. The TDGL equations (in Gaussian units) are given by:
\begin{widetext}
\begin{align}
    \Gamma\left( \frac{\partial \psi}{\partial t} + \frac{i e_s \phi}{\hbar}\psi \right) + \frac{1}{2m_s}\left(-i\hbar\nabla - \frac{e_s}{c}\mathbf{A}\right)^2\psi + \alpha \psi + \beta |\psi|^2\psi = 0 \label{psi_eq_units} \\
    \frac{4\pi \sigma_n}{c}\left( \frac{1}{c}\frac{\partial \mathbf{A}}{\partial t} + \nabla\phi \right) + \curl{\curl{\mathbf{A}}} - \frac{2\pi i e_s \hbar}{m_s c}\left( \psi^*\nabla\psi - \psi\nabla\psi^*\right) + \frac{4\pi e^2_s}{m_s c^2}|\psi|^2\mathbf{A} = 0. \label{j_eq_units}
\end{align}
\end{widetext}
These equations are solved for the complex superconducting order parameter, $\psi$, and the magnetic vector potential, $\mathbf{A}$. The magnitude squared of $\psi$ is proportional to the density of superconducting electrons. The parameters $\alpha$ and $\beta$ are phenomenological, and were originally introduced as coefficients of the series expansion of the Ginzburg-Landau free energy. Additionally, $\phi$ is the scalar potential; $\sigma_n$ is the normal electron conductivity; $\Gamma$ is the phenomenological relaxation rate of $\psi$. Furthermore, $e_s = 2e$ and $m_s = 2m_e$ represent the total charge and total effective mass of a Cooper pair, respectively. The TDGL equations are subject to boundary conditions
\begin{align}
    \left( i\hbar\nabla\psi + \frac{e_s}{c}\mathbf{A}\psi \right)\cdot \mathbf{n} = 0 \label{current_bc} \\ 
    \left(\curl{\mathbf{A}}\right)\times \mathbf{n} = \mathbf{H}_a\times \mathbf{n} \label{H_bc} \\ 
    \left( \nabla\phi + \frac{1}{c}\frac{\partial \mathbf{A}}{\partial t} \right)\cdot \mathbf{n} = 0, \label{E_bc}
\end{align}
where $\mathbf{n}$ is the outward normal vector to the boundary surface and $\mathbf{H}_a$ is the applied magnetic field. Eq. \ref{current_bc} ensures no current flows out of the superconducting domain, and noting that $E = -\nabla\phi - \frac{1}{c}\frac{\partial \mathbf{A}}{\partial t}$, Eqs. \ref{H_bc} and \ref{E_bc} are electromagnetic interface conditions with an applied magnetic field.

The parameters $\alpha$, $\beta$, and $\Gamma$ were originally introduced into the theory as phenomenological, temperature-dependent constants \cite{duGL}. It is worth noting that $\alpha < 0$ corresponds to the superconducting state whereas $\alpha \geq 0$ corresponds to the normal state; $\beta$ is strictly positive regardless of the system's state.  The TDGL equations can also be derived from microscopic theory using the time-dependent Gor'Kov equations \cite{GorKov_TDGL}. A useful consequence of this derivation is that it allows the TDGL parameters to be directly related to experimentally observable properties of the material in question. The material dependencies are given by Ref. \citenum{kopnin2001theory}:
\begin{align}
    \alpha(\nu(0),T_c,T) &= -\nu(0)\left(\frac{1-\frac{T^2}{T_c^2}}{1+\frac{T^2}{T_c^2}}\right) \label{alpha_eq} \\
    &\approx -\nu(0)\left(1-\frac{T}{T_c}\right) \nonumber \\
    \beta(\nu(0),T_c,T) &= \frac{7\zeta(3)\nu(0)}{8\pi^2T_c^2}\left(\frac{1}{1+\frac{T^2}{T_c^2}}\right)^2 \label{beta_eq} \\ &\approx \frac{7\zeta(3)\nu(0)}{8\pi^2T_c^2} \nonumber \\
    \Gamma(\nu(0),T_c) &= \frac{\nu(0)\pi\hbar}{8 T_c}, \label{gamma_eq}
\end{align}
where $\nu(0)$ is the density of states at the Fermi-level, $T_c$ is the critical temperature, $T$ is the temperature, and $\zeta(x)$ is the Riemann zeta function. Additionally, the effective mass can also be expressed in terms of these same quantities in addition to the Fermi velocity, $v_f$, and electron mean free path, $\ell$:
\begin{equation}
    m_s = \frac{12\hbar T_c}{\pi \nu(0)v_f \ell}.\label{m_eff_eq}
\end{equation}
Equation \ref{m_eff_eq} gives the effective mass under the dirty limit.

A major contribution of this paper is to introduce a framework for modeling sample-specific features of superconducting materials by connecting \textit{ab initio} calculations of the material's properties to experimental characterizations of the material. Eqs. \ref{alpha_eq}-\ref{m_eff_eq} determine how the parameters of TDGL depend on the underlying material properties. We allow these parameters to vary spatially to capture the sample-specific features observed from experimental characterizations. When converting these constant coefficient into a spatially-varying ones, it is important to consider how this may alter the TDGL equations due to potential alterations to the functional derivatives of the free energy. In the case of $\alpha$, $\beta$, and $\Gamma$, they do not appear on terms of the free energy with spatial gradients. However, $m_s$ does appear in terms including gradients, and so when including spatially-varying effective mass, Equation \ref{psi_eq_units} should be augmented with an additional term:
\begin{equation*}
    \frac{i\hbar}{2} \nabla\frac{1}{m_s}\cdot(i\hbar \nabla + \frac{e}{c}\mathbf{A})\psi
\end{equation*}

Additionally, when solving the TDGL equations numerically, it is standard to normalize all the parameters of the model in order to obtain dimensionless quantities. In order to satisfy the above requirements, the steps are as follows: Pick a reference value for each of $\alpha$, $\beta$, $\Gamma$, $m_s$, and $\sigma_{n}$, these will most often just be the corresponding values for the bulk material. Label these reference values $\alpha_0$, $\beta_0$, $\Gamma_0$, $m_{0}$, and $\sigma_{n0}$. Define dimensionless spatially varying functions, $a(\mathbf{r})$, $b(\mathbf{r})$, $\gamma(\mathbf{r})$, $\mu(\mathbf{r})$, and $s(\mathbf{r})$, relative to their reference values. Apply the following transformations to Eqs. \ref{psi_eq_units} and \ref{j_eq_units}:
\begin{align}
    \alpha &\longrightarrow \alpha_0 a(\mathbf{r}) \label{alpha_scale} \\
    \beta &\longrightarrow \beta_0 b(\mathbf{r}) \label{beta_scale} \\
    \Gamma &\longrightarrow \Gamma_0 \gamma(\mathbf{r}) \label{gamma_scale} \\
    m_s &\longrightarrow m_0 \mu(\mathbf{r}) \label{m_eff_scale} \\
    \sigma_n &\longrightarrow \sigma_{n0} s(\mathbf{r}). \label{sigma_scale}
\end{align}
Then proceed with standard nondimensionalization procedures for TDGL (see the Appendix for more details). The advantage of this approach is that the nondimensionalization procedures, when used on these transformed equations, leave behind only the dimensionless functions which capture the spatial variation of the sample material in natural units. The resulting equations are
\begin{widetext}
\begin{align}
    \gamma\left(\frac{\partial \psi}{\partial t} + i \kappa_0\phi\psi\right) + \frac{1}{\mu}\left(\frac{-i}{\kappa_0}\nabla - \mathbf{A}\right)^2\psi + \frac{1}{\kappa_0}\nabla\frac{1}{\mu}\cdot\left(\frac{1}{\kappa_0}\nabla - i\mathbf{A}\right)\psi - a\psi + b|\psi|^2\psi = 0 \label{psi_eq}
    \\
    \frac{s}{u_0}\left(\frac{\partial \mathbf{A}}{\partial t} + \nabla\phi\right) + \curl{\curl{\mathbf{A}}} + \frac{i}{2\kappa_0 \mu}\left( \psi^* \nabla\psi - \psi\nabla\psi^*\right) + \frac{1}{\mu}|\psi|^2\mathbf{A} = 0, \label{current_eq}
\end{align}
\end{widetext}
where $\kappa_0 = \frac{\lambda_0}{\xi_0}$ is the Ginzburg-Landau parameter of the reference material. The quantity $\lambda_0 = \sqrt{\frac{m_0 c^2 \beta_0}{4\pi e^2_s |\alpha_0|}}$ is the penetration depth of the reference material, and $\xi_0 = \sqrt{\frac{\hbar^2}{2m_0|\alpha_0|}}$ is its coherence length. The parameter $u_0 = \frac{\tau_{\psi_0}}{\tau_{j_0}}$ is a ratio of characteristic time scales in the reference material, where $\tau_{\psi_0} = \frac{\Gamma_0}{|\alpha_0|}$ is the characteristic relaxation time of $\psi$ in the reference material and $\tau_{j_0} = \frac{\sigma_{n0}m_0\beta_0}{e_s^2|\alpha_0|}$ is the characteristic relaxation time of the current. We have also inserted a minus in front of $a$, which is just a convention to make positive values of $a$ correspond to the superconducting state (Note: this is the opposite of how $\alpha$ is usually interpreted within Ginzburg-Landau theory, however it is standard to make this change when performing nondimensionalization). The boundary conditions become:
\begin{align}
    \left( \frac{i}{\kappa_0}\nabla\psi + \mathbf{A}\psi \right)\cdot n = 0 \\
    \left(\curl{\mathbf{A}}\right)\times n = \mathbf{H}_a\times n
    \\
    \left( \nabla\phi + \frac{\partial \mathbf{A}}{\partial t} \right)\cdot n = 0.
\end{align}
It should be noted that $\gamma(\mathbf{r})$ and $s(\mathbf{r})$ allow the local characteristic time relaxations to vary, which only affects the dynamics of the solutions. Where the time dynamics are relevant these parameters cannot be ignored; however, in many cases, the primary interest of TDGL calculations is in determining the energetic stability and critical fields, such as the superheating field. In these cases, the time dynamics are not relevant and $\gamma$ and $s$ can be safely set to arbitrary constant values, such as unity.

\subsection{Determining Spatial Variation of TDGL Parameters}

In the previous section, we have shown how to introduce spatial variation to the TDGL parameters. The process of calculating the values of $a(\mathbf{r})$, $b(\mathbf{r})$, $\gamma(\mathbf{r})$, and $\mu(\mathbf{r})$ is where we bring in experiment and \textit{ab initio} theory. From Eqs. \ref{alpha_eq}-\ref{gamma_eq}, we know these parameters mostly depend on well-defined microscopic quantities, namely $\nu(0)$, $T_c$, and $v_f$. These quantities can be calculated using density-functional theory (DFT). Local densities of states and Fermi velocities are straightforward to compute in DFT, providing local values for $\nu(0)$ and $v_f$. Superconducting quantities such as $T_c$ are calculated by applying Eliashberg theory within a DFT framework and directly calculating electron-phonon coupling from first principles. Experiment can then give information about the compositions of sample materials, and DFT calculations can determine the $\nu(0)$, $v_f$, and $T_c$ associated with these compositions. Using these values in addition to estimates of the electron mean free path (which can be derived from DFT or can come from experimental characterizations), $a(\mathbf{r})$, $b(\mathbf{r})$, $\gamma(\mathbf{r})$,and $\mu(\mathbf{r})$ are calculated from Eqs. \ref{alpha_eq}-\ref{m_eff_eq}, and the material geometries from the experimental results determine the spatial variation.

In this paper, we will demonstrate our framework on Sn-deficient islands. These defects have been studied extensively \cite{Trenikhina_Patchy_Grains_Sn_islands,Lee_Patchy_Grains_Sn_islands,Becker_Sn_islands,Viklund_Sn_Islands} so it is straightforward to find estimates of the general size of these islands and their material compositions in the literature. DFT has also been used to calculate $\nu(0)$, $v_f$, and $T_c$ for Nb-Sn systems of different concentrations. In this paper we use the DFT values from Ref. \citenum{Sitaraman_Nb3Sn} to construct $a(\mathbf{r})$, $b(\mathbf{r})$, $\gamma(\mathbf{r})$, and $\mu(\mathbf{r})$ for islands of different Sn concentration. This paper will primarily focus on the results of our TDGL calculations, for further information regarding the details and results from the DFT calculations used in this paper, we refer the interested reader to Refs. \citenum{Kelley_GBs} and \citenum{Sitaraman_Nb3Sn}.

\subsection{Simulation Geometry and Numerical Implementation} \label{SimulationGeometry}

The simulations in this paper were solved using a finite element formulation of the TDGL equations proposed by Gao \cite{gao3D}, under the temporal gauge, which sets the scalar potential $\phi = 0$ (see Du \cite{duGauges} for a more detailed overview of gauge choices in TDGL). All computations were carried out using the open-source finite element software FEniCS \cite{alnaes2015fenics}.

\begin{figure}
    \includegraphics[width=0.95\columnwidth]{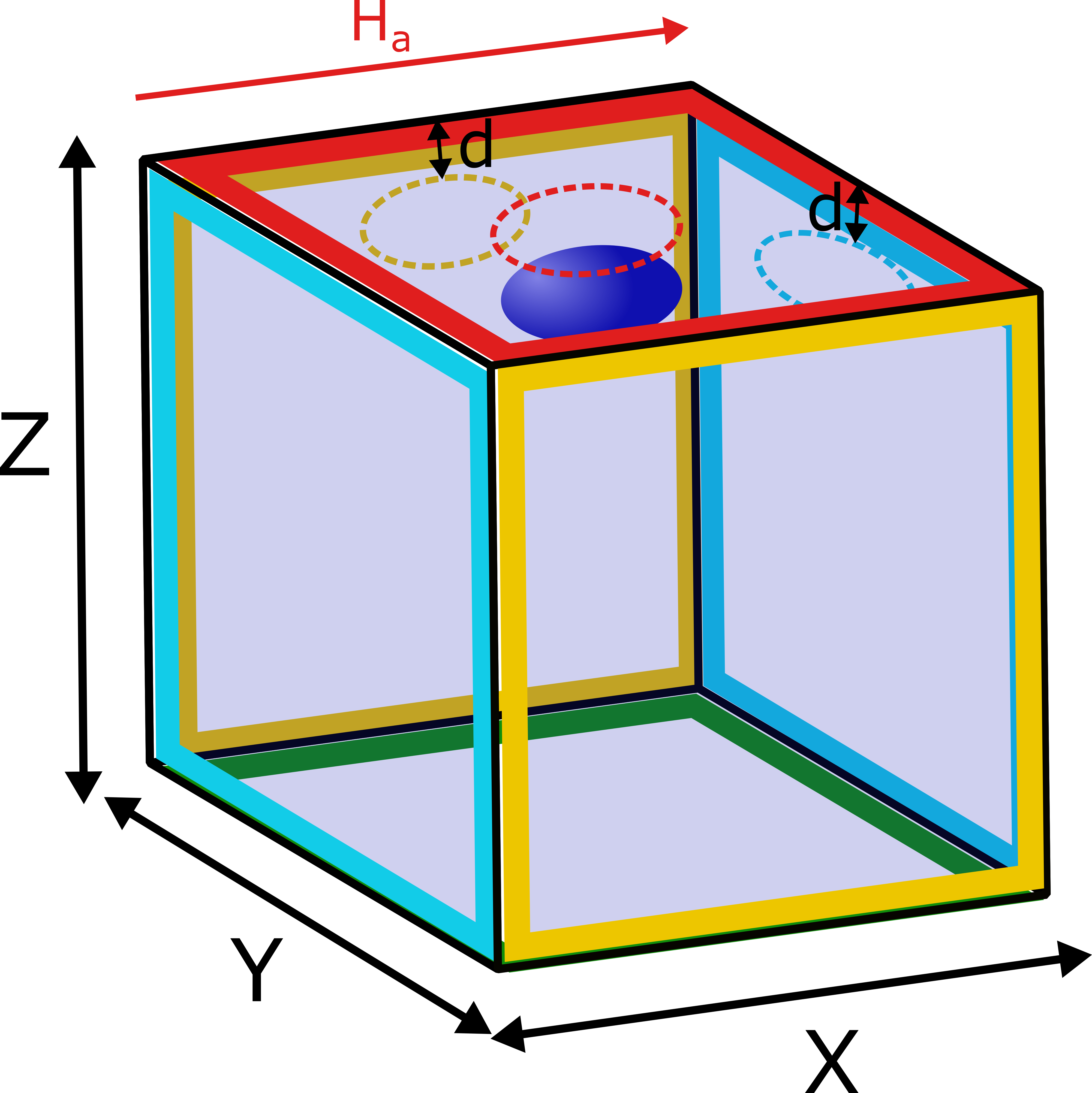}
    \caption{\justifying \textbf{Schematic of Simulation Geometry.} Like-colored surfaces (yellow and light blue outlines) have periodic boundary conditions. The red and green surfaces indicate the faces where external magnetic field is applied; in simulations, a field is applied to the red surface in the $x$-direction (indicated by the red arrow), with no field applied on the green surface. The Sn-deficient island is shown as a dark blue ellipsoid, with dotted lines indicating its projections into the $xy$, $xz$, and $yz$ planes. The distance $d$ is measured from the surface to the outer edge of the ellipsoid.}
    \label{fig:3DAlpha}
\end{figure}

The simulation domain consists of a rectangular cuboid with periodic boundary conditions applied in the $x$- and $y$-directions, as indicated by the yellow and light blue highlights in Fig.~\ref{fig:3DAlpha}. An external magnetic field of constant magnitude $H_a$ is applied in the $x$-direction on the red surface, while no field is applied on the opposing green surface. For simulations that include Sn-deficient islands, the island geometry is also shown in Fig.~\ref{fig:3DAlpha}. These defects are modeled as ellipsoids with equal radii in the $x$- and $y$-directions and a $z$-radius equal to half of the $x$-radius. The island's distance from the surface, denoted by $d$, is measured from the surface to the outer edge of the ellipsoid.

All meshes were generated using the open-source mesh generation software Gmsh \cite{geuzaine2009gmsh}, with cubic tetrahedral elements. For simulations involving Sn-deficient islands, the OpenCASCADE geometry kernel in Gmsh was used to ensure that the mesh conforms to the ellipsoidal shape of the defects.

\subsection{Dissipation in TDGL}

When simulating SRF materials, dissipation is often a physical quantity of interest. Under TDGL, a formula for dissipation can be derived by considering the time derivative of the free energy and the free energy current flux density. A more detailed derivation is found in Ref. \citenum{kopnin2001theory}, but we quote the final result here:
\begin{equation}
    D = 2\Gamma\left|\left(\frac{\partial \psi}{\partial t} + \frac{ie_s\phi\psi}{\hbar}\right)\right|^2 + \sigma_n \mathbf{E}^2. \label{dissipation_eq}
\end{equation}
This quantity is a power density, with the first term corresponding to the superconducting dissipation arising from the relaxation of the order parameter. The second term is the dissipation of normal currents which are largest near the surface where magnetic field can still appreciably penetrate.

It is worth considering how this expression for the dissipation density is related to existing theories of RF power loss and surface resistance. The first term in Eq. \ref{dissipation_eq} is associated with the dissipation due to the rate of change of $\psi$. This term is typically small, except in the vortex state where it becomes the dominant source of dissipation. A dissipation of this form is similar to that proposed by Tinkham \cite{tinkham_flux_flow}, who suggested the vortex dissipation should be proportional to $\left(\frac{\partial \psi}{\partial t}\right)^2$. The additional term within the parenthesis in Eq. \ref{dissipation_eq} is a result of the gauge invariance of TDGL. 

The second term in Eq. \ref{dissipation_eq} can be directly related to the phenomenological ``two-fluid model," which was first proposed by Gorter and Casimir \cite{gortercasimir_2fluid} in 1934 and was applied by London \cite{London_2fluid} later that year to calculate the power loss of a superconductor. The two-fluid model approximates the electrons of the system as consisting of two non-interacting `fluids': the superconducting electrons, in the form of cooper pairs which carry lossless supercurrent, and the normal electrons, which exist as thermally excited quasiparticles that produce typical dissipative currents. Under the two-fluid model, the normal fluid losses produce dissipation of the form \cite{halbritter_surface_resistance,turneaure_mattis-bardeen}
\begin{equation*}
    P \propto \sigma_n E^2,
\end{equation*}
which is identical to the second term of Eq. \ref{dissipation_eq}. For AC applied currents, the electric field is proportional to the frequency and magnetic field, $E \propto \omega H$, meaning that overall the power loss will be of the form 
\begin{equation} \label{2FM_dissipation}
    P \sim \omega^2 H^2. 
\end{equation}
It is also well known that within RF cavities, the power loss is given by
\begin{equation} \label{RF_dissipation}
    P = \frac{1}{2}\oint_A R_s |H(r)|^2 dA,
\end{equation}
where $R_s$ is the surface resistance of the cavity walls and $A$ is the surface area. Comparing Eqs. \ref{2FM_dissipation} and \ref{RF_dissipation}, we have that the second term of Eq. \ref{dissipation_eq} directly leads to a surface resistance proportional to the square of the frequency, $R_s \propto \omega^2$. In the late 1950s, Mattis and Bardeen \cite{mattis-bardeen} as well as Abrikosov, Gor'Kov, and Khalatnikov \cite{abrikosovgorkovkhalatnicov} independently derived the now well-known ``BCS resistance."
Under the low frequency and low field limit, the BCS resistance reduces to the form \cite{gurevich_surface_resistance,Gurevich_SRF_DOS}
\begin{equation}
    R_{bcs} \simeq \frac{\omega^2}{T}e^{\frac{-\Delta}{k_BT}} \label{BCS_resistance}
\end{equation}
where $\Delta = 1.76k_BT_c$ is the superconducting energy gap \cite{kopnin2001theory}. We thus see that our calculated expression for the surface resistance has the same $\omega^2$ frequency dependence of the BCS prediction in the low frequency and field limit, though the quantitative values may differ.

Given this connection to surface resistance, we now consider how our dissipation estimates can be used to calculate the cavity quality factor, $Q$, a standard figure of merit in SRF cavity performance.

\subsection{Estimating Cavity Quality Factor} \label{Estimating_Quality_Factor}

While TDGL allows us to estimate dissipation, it is important to emphasize that both the dissipation calculations and the quality factor estimates that follow lie well outside the regime of quantitative validity for the theory. In particular, TDGL is strictly valid only near $T_c$ in the gapless limit, and its predictions for dissipation under RF-like dynamic fields at low temperatures should be interpreted with caution. Despite this, we believe the calculations presented here remain qualitatively valuable. They provide a means of linking mesoscopic-scale simulations to macroscopic cavity performance metrics, and enable relative comparisons between different material configurations that may inform experimental priorities.

A particularly relevant quantity that can be estimated from the dissipation is the cavity quality factor, $Q$, which is given by
\begin{equation}
    Q = \frac{2\pi E}{\Delta E}. \label{quality_factor_eq}
\end{equation}
Here, $E$ is the energy stored in the cavity, and $\Delta E$ is the energy dissipated in the cavity walls during each RF cycle. It is common to express the quality factor in terms of the surface resistance as
\begin{equation} \label{reduced_quality_factor_eq}
    Q = \frac{G}{R_s},
\end{equation}
where $R_s$ is the cavity surface resistance and $G$ is a geometric factor that depends only on quantities which are determined by the cavity geometry. For a typical 1.3 GHz 9-cell Nb TESLA cavity, $G = 270$ $\Omega$ \cite{Nb_SRF_Cavity_Info}. The surface resistance is given by
\begin{equation}
    R_s = \frac{\mu_0 \omega \lambda^3}{\Tilde{H_a^2} L_x L_y}\left(I_\psi + \frac{\omega}{\tilde{\omega}} \sigma_n \mu_0 \lambda^2 I_A\right), \label{surface_resistance_eq}
\end{equation}
where $\mu_0$ is the permeability of free space, $\omega$ is the cavity frequency, $\lambda$ is the penetration depth, $\Tilde{H_a}$ is the maximum applied magnetic field value in simulation units, $L_x$ and $L_y$ are the size of a simulation domain in the X and Y directions respectively, $\sigma_n$ is the normal conductivity, and $\tilde{\omega}$ is the frequency of the applied field in simulation units. $I_\psi$ and $I_A$ are integrals over the squared time derivatives of $\psi$ and $\mathbf{A}$:
\begin{align}
     I_\psi &\equiv \int d\Tilde{t} \int d\Tilde{x} \int d\Tilde{y} \int d\Tilde{z} \left|\frac{\partial\Tilde{\psi}}{\partial \Tilde{t}}\right|^2 \\
     I_A &\equiv \int d\Tilde{t} \int d\Tilde{x} \int d\Tilde{y} \int d\Tilde{z} \left(\frac{\partial\Tilde{\mathbf{A}}}{\partial \Tilde{t}}\right)^2
\end{align}
where the tilde variables denote ones which are in simulation units. A much more detailed derivation of these equations can be found in the Appendix.

The $\sigma_n$ appearing in Eq. \ref{surface_resistance_eq} refers specifically to the conductivity of the normal quasiparticles. In general, calculating this quantity from first principles is a complex task that depends sensitively on microscopic material properties, and is largely orthogonal to the rest of the quality factor calculation. For this reason, we choose to treat $\sigma_n$ as a free parameter in our model, which allows us to explore how dissipation and quality factor vary across a range of plausible conductivity values without relying on potentially oversimplified assumptions. For completeness, we outline a possible approximate derivation of $\sigma_n$ based on the Drude model in the Appendix, though we emphasize that this is intended primarily as a qualitative reference.

In the derivation of Equation \ref{surface_resistance_eq}, it is assumed that the cavity surface is partitioned into small fractional areas, and the dissipated energy is calculated over one of these areas, and then multiplied by the total number of them. If some defect is present in the simulation domain, it would mean that a $Q$ calculation based purely on that value would implicitly assume that such a defect is uniformly distributed over the surface of the cavity. In Equation \ref{surface_resistance_eq}, the only thing that changes when simulating a different material or different defect is the value of the quantity in parenthesis, every other part of the process for calculating $Q$ remains the same. This means that we can calculate a more realistic cavity surface resistance by calculating $R_s$ with Equation \ref{surface_resistance_eq} for a few different simulations, and then taking a weighted average of these values. Let $R_i$ be the surface resistance of the $i$th simulation, and let $p_i$ be the percentage of the fractional areas partitioning the cavity surface which are represented by this simulation. Then
\begin{equation}
    R_{tot} = \sum_i p_i R_i,\label{total_resistance_eq}
\end{equation}
where $\sum_i p_i = 1$. The simplest application for Equation \ref{total_resistance_eq} is to perform two simulations, one baseline simulation with no defects or material inhomogeneity, and then another simulation containing some defect of interest. Choosing a value $p$ to represent the percentage of fractional areas containing the defect and then following through with the rest of the quality factor calculation provides a simple way to estimate $Q$ for different average defect densities by simply changing the value of $p$.

The value of Q calculated from TDGL outputs will typically be underestimated at low field. This is because of the assumption of gapless superconductivity, which results in higher surface resistances than is predicted with the BCS surface resistance (Equation \ref{BCS_resistance}). Despite this, our approach still often predicts quality factors within an order of magnitude of the experimental values. Additionally, the relative behavior of Q at different applied fields, especially when averaging the impact of multiple kinds of defects as described above, qualitatively captures effects such as high field Q-slope. Nonetheless, we emphasize that this quality factor calculation is best treated as a qualitative tool; for quantitatively accurate predictions, more rigorous superconductivity theories should be used.

\section{Validation Study}\label{Results}

\subsection{Sn-deficient Islands in Nb\textsubscript{3}Sn} \label{section_sn_islands}

There has been a considerable amount of work both experimentally and theoretically understanding the various defects present in Nb\textsubscript{3}Sn \cite{Posen_Valles_Liepe_field_limits,  Trenikhina_Patchy_Grains_Sn_islands, Lee_Patchy_Grains_Sn_islands, Pudasaini_Patchy_Grains, Lee_Sn_Segregated_GBs, Oh_Diffusion_GBs, Carlson_GBs, Becker_Sn_islands, Viklund_Sn_Islands}. For the purposes of this paper, we will focus on Sn-deficient islands, small regions within Nb\textsubscript{3}Sn grains that have lower than expected Sn concentrations. To demonstrate the use of our sample specific TDGL framework, we will estimate the potential impact to SRF performance by determining the vortex penetration field, $H_{vort}$. This field represents the lowest applied field at which a vortex nucleates and generalizes the concept of the superheating field to surfaces containing defects. In the limit of a uniform, flat surface, $H_{vort}$ is equal to the superheating field.

In order to simulate Sn-deficient islands, we require estimates of the material properties within these regions. Such estimates are provided in Ref.~\citenum{Sitaraman_Nb3Sn}, which reports values of $T_c$ and $\nu(0)$ for Nb-Sn systems at 18.7, 20.8, and 23.4 at.\% Sn. Our $v_f$ estimates for these compositions were obtained via private communication \cite{private2025}. We first choose a Sn concentration of 18.7 at.\% for our islands. Using the corresponding values of $T_c$ and $\nu(0)$, we construct $a(\mathbf{r})$, $b(\mathbf{r})$, and $\gamma(\mathbf{r})$; the values of these quantities in the bulk as well as in the island are reported in Table~\ref{tab:TDGL_Parameter_Values}. The simulation geometry is the same as depicted in Fig.~\ref{fig:3DAlpha}, with the domain a cube of side length $10\lambda$. The applied field for these simulations is held constant.
\begin{table}[]
    \centering
    \begin{tabular}{|p{4.0cm}||p{1.0cm}|p{1.0cm}| }
    \hline
    \multicolumn{3}{|c|}{TDGL Parameter and Material Values} \\
    \hline
    Quantity & Bulk Value & Island Value \\
    \hline
    $T_c$ (K) & 18 & 6 \\
    $\nu(0)$(states/(eV*nm$^3$)) & 101.33 & 40.53\\
    $v_f$ ($10^5$ m/s) & 1.25 & 1.4\\
    $\ell$ (nm) & 5.5 & 2.25\\
    $a$ (unitless) & 1 & 0.15\\
    $b$ (unitless) & 1 & 1.8\\
    $\gamma$ (unitless) & 1 & 1.2 \\
    $s$ (unitless) & 1 & 1 \\
    $\mu$ (unitless) & 1 & 1.79\\
    \hline
    \end{tabular}
    \caption{\justifying \textbf{A summary of the TDGL parameter and material values used in the results of this paper.} All TDGL parameters were calculated from Equations \ref{alpha_eq}-\ref{gamma_eq} using the higher order temperature dependence and normalized with respect to the bulk value. The $T_c$ and $\nu(0)$ values were taken from Ref. \citenum{Sitaraman_Nb3Sn} assuming a Sn concentration of 18\% for the island. The $v_f$ values came from private correspondence \cite{private2025}, and the bulk value of $\ell$ was taken from Ref. \citenum{Mentink_mfp_ref}, we set the island value to half of the bulk value. The normal electron conductivity, $s$, was assumed to remain constant throughout the simulation domain.}
    \label{tab:TDGL_Parameter_Values}
\end{table}

\begin{figure}
  \begin{subfigure}{\columnwidth}
    \centerline{\includegraphics[width=0.675\columnwidth]{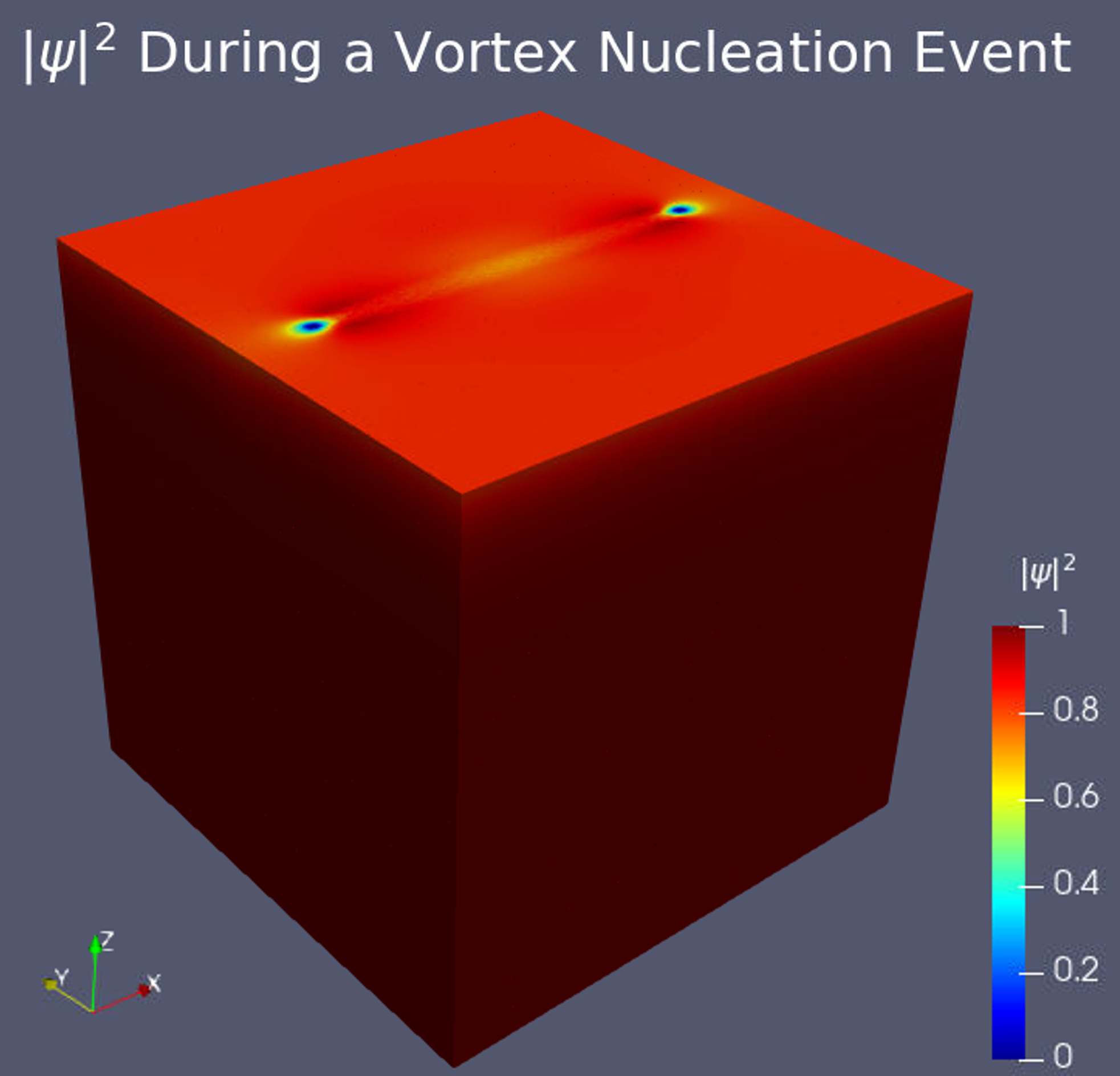}}
    \caption{A plot of $|\psi|^2$ during a vortex nucleation event.}
  \end{subfigure}
  \begin{subfigure}{\columnwidth}
    \centerline{\includegraphics[width=0.675\columnwidth]{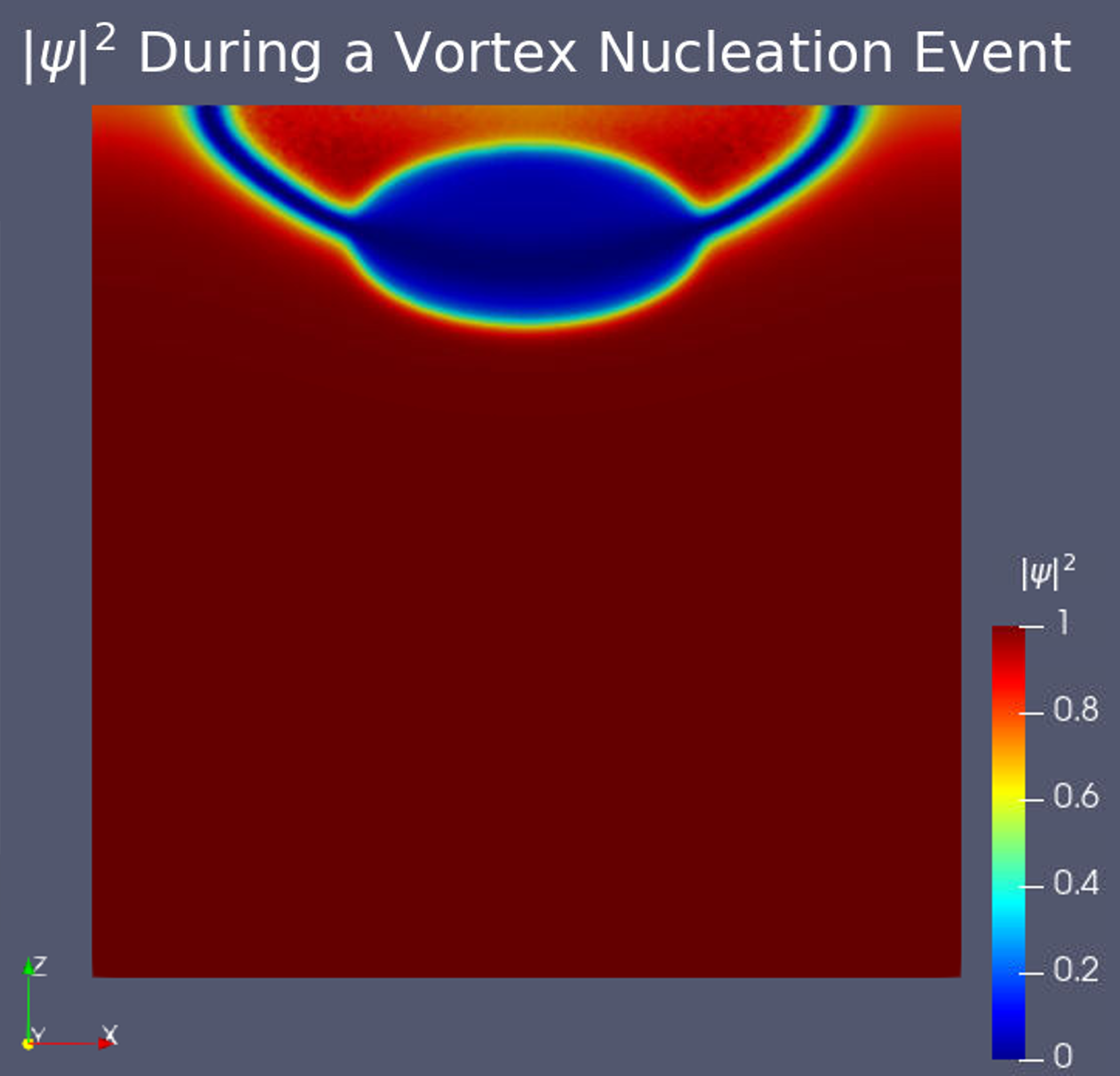}}
    \caption{A 2D slice of $|\psi|^2$ in the XZ plane at $\frac{y}{\lambda} = 5$}
  \end{subfigure}
  \begin{subfigure}{\columnwidth}
    \centerline{\includegraphics[width=0.675\columnwidth]{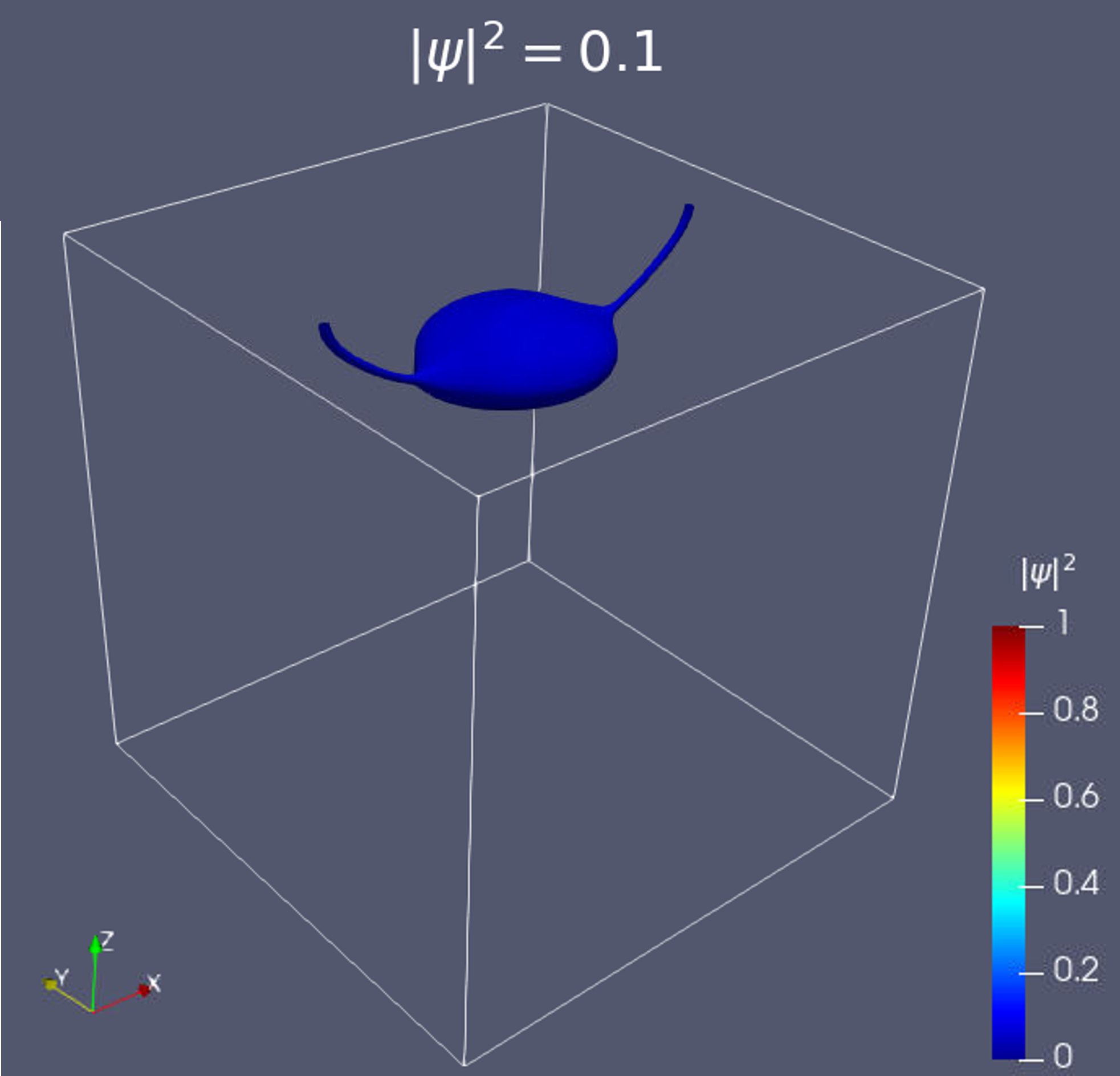}}
    \caption{The $|\psi|^2 = 0.1$ isosurface.}
  \end{subfigure}
  \caption{\justifying \label{fig:Sn_Islands_Sim_Example}  \textbf{Example of a Sn-deficient Island Simulation.} The domain is a cube with side length $10\lambda$, this particular simulation is for an island $0.5\lambda$ from the surface, with a X and Y radius of $2\lambda$ and a Z radius of $1\lambda$. The applied field is $\frac{H_a}{\sqrt{2}H_c} = 0.38$, which is the value of $H_{vort}$ for this island size and distance from the surface. (a) depicts a volume plot of $|\psi|^2$ over the whole domain during a vortex nucleation event. (b) is a 2D slice in the XZ plane at $\frac{y}{\lambda} = 5$. (c) is the $|\psi|^2 = 0.1$ isosurface, which shows the shape of the vortex as well as the island that induced the nucleation.}
\end{figure}

\begin{figure}
  \includegraphics[width=1.0\columnwidth]{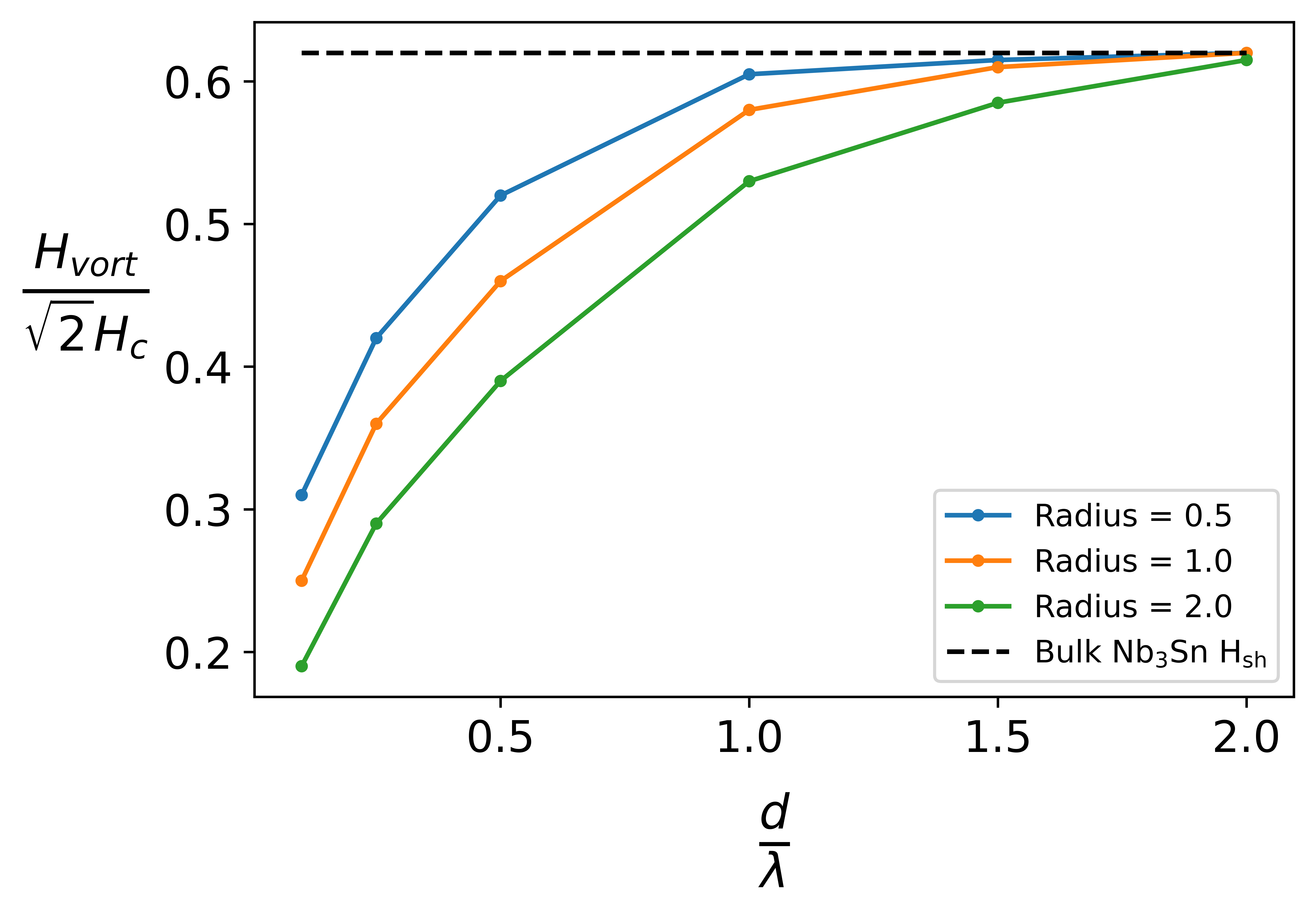}
  \caption{\justifying \label{fig:Sn_Islands_Hsh}
    \textbf{Vortex Penetration Field Versus Distance from Surface for Different Island Size.}
    We plot our calculated vortex penetration field estimates, varying the distance of the Sn deficient island from the superconductor surface. The field values are reported in units of $\sqrt{2}H_c$ and the distances are in units of penetration depths. We did this for 3 different islands, with x radii of 0.5, 1, and 2 penetration depths. The black dotted line denotes the superheating field of bulk Nb\textsubscript{3}Sn. Smaller volume islands have a diminishing impact on $H_{vort}$.  
  }
\end{figure}

\begin{figure}
  \includegraphics[width=1.0\columnwidth]{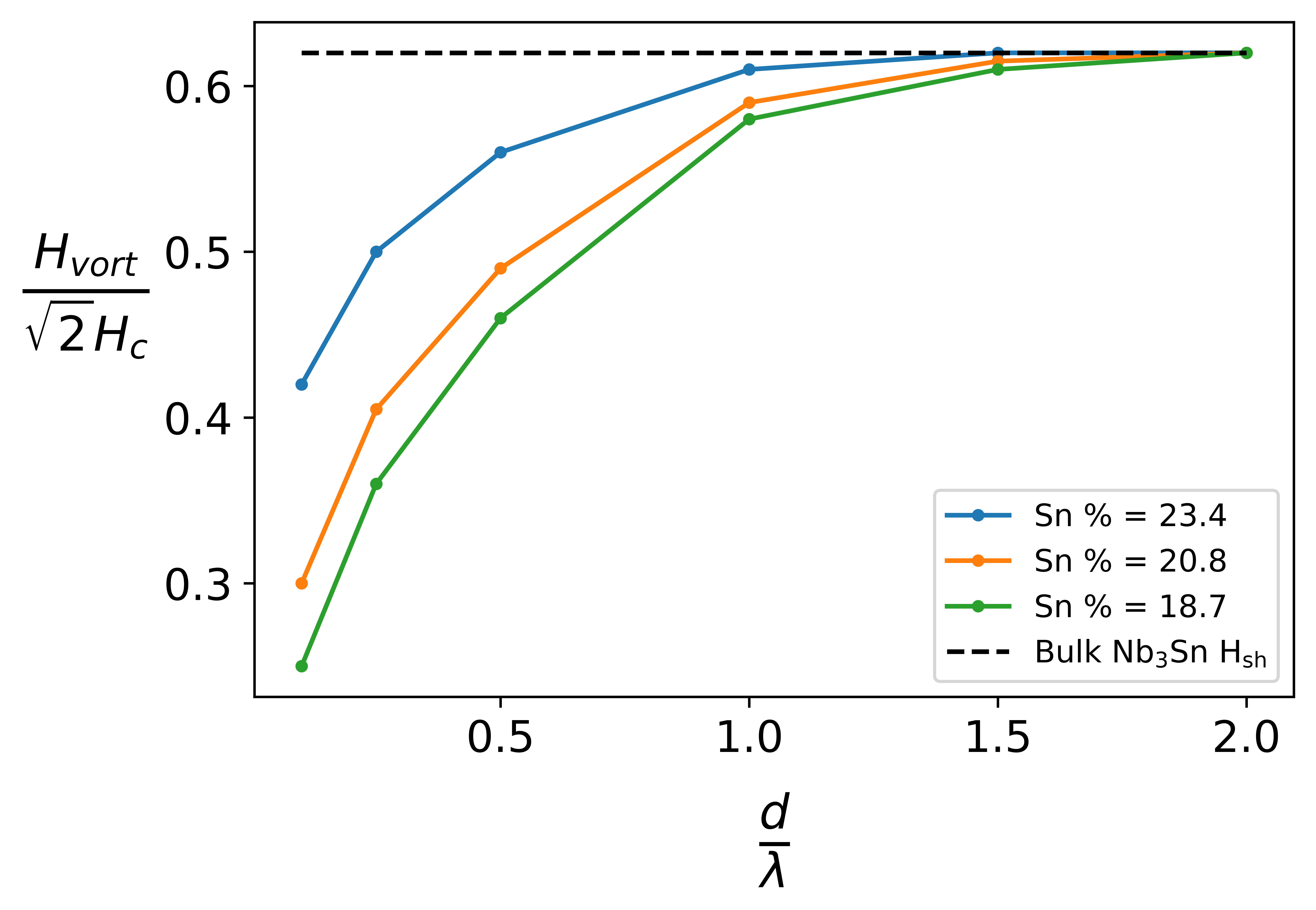}
  \caption{\justifying \label{fig:Sn_Islands_Hsh_Sn_Pcnt}
    \textbf{Vortex Penetration Field Versus Distance from Surface for Different Island Sn \%.}
    We plot our calculated vortex penetration field estimates, varying the distance of the Sn deficient island from the superconductor surface. The field values are reported in units of $\sqrt{2}H_c$ and the distances are in units of penetration depths. We did this for 3 different island Sn percentages, $18.7\%$, $20.8\%$, and $23.4\%$. The black dotted line denotes the superheating field of bulk Nb\textsubscript{3}Sn.  
  }
\end{figure}

Figure \ref{fig:Sn_Islands_Sim_Example} depicts vortex nucleation occuring at $H_{vort}$ for a particular island. We find that varying the distance of the island from the surface between $0.1$ and $2$ penetration depths (i.e. between $\sim$10-200 nm for Nb\textsubscript{3}Sn), we observe a reduction in $H_{vort}$ by as much as $\sim$60\% below the bulk value of the superheating field for islands very near the surface, as shown in Fig. \ref{fig:Sn_Islands_Hsh}. We did this for several different island sizes, and found that the severity of this drop in $H_{vort}$ increases with increasing island size. As such, we conclude that large islands within 200 nm of the cavity surface are a potential cause for concern when it comes to SRF performance, particularly ones within 1-2 penetration depths of the surface.  Figure \ref{fig:Sn_Islands_Hsh_Sn_Pcnt} shows how these effects change with respect to the Sn at.\% inside each island. We see that as the Sn deficiency becomes weaker, so does the impact on vortex penetration, though it is worth noting that even an island with 23.4\% Sn (only 1.6\% off a perfect stoichiometric ratio) has a $\sim$30\% reduction in $H_{vort}$ for islands near the surface. Currently, Nb\textsubscript{3}Sn-coated SRF cavities are only achieving at most $\sim$25\% of their theoretical maximum accelerating fields. There are likely many factors which contribute to this outcome, but these results indicate that Sn-deficient islands are among the defects which contribute to SRF performance degradations.

\subsection{Sources of Dissipation in Nb SRF Cavities}

In the previous section, we focused on steady-state properties of TDGL in the dirty limit, where the theory is known to have broader quantitative validity. We now extend this framework to examine the two dissipation terms that naturally arise in TDGL simulations. While these calculations fall outside the strict regime of TDGL's quantitative accuracy, particularly at the low temperatures and high frequencies relevant for SRF cavities, they remain \textit{qualitatively} informative. In particular, they allow us to interpret distinct contributions to RF dissipation in a way that parallels the conventional decomposition of surface resistance into BCS and residual components. More broadly, they provide a means of linking mesoscopic TDGL behavior to macroscopic cavity losses, and of understanding how steady-state properties, such as $H_{vort}$ for individual defects, can collectively influence large-scale phenomena like the high-field Q-slope.

It is standard in SRF literature to decompose surface resistance into two components: the BCS resistance, which arises from thermally excited quasiparticles, and a more poorly understood residual resistance, which dominates at low temperatures and is often attributed to trapped flux, impurities, or surface defects. In our TDGL framework, these two types of dissipation emerge naturally. Specifically, in Eq.~\ref{surface_resistance_eq}, the term proportional to $\sigma_n$ (involving $I_A$) captures dissipation due to normal current response, and is the TDGL analog of the BCS resistance. The second term ($I_\psi$), which arises from relaxation of the superconducting order parameter, reflects dissipation mechanisms not captured by traditional two-fluid or BCS models and can be viewed as a proxy for residual resistance in our simulations. Studying the behavior and interplay of these two terms as a function of applied field and conductivity provides insight into which mechanisms dominate under different conditions.

To establish a baseline for comparison, we simulate a uniform Nb domain subjected to a sinusoidal applied field with period $2000\tau_\psi$. This choice is motivated by the fact that, for Nb at 2 K, we estimate $\tau_\psi \approx 3.7 \times 10^{-13}$ s using Eqs.~\ref{alpha_eq} and \ref{gamma_eq} and a value of $\nu(0) = 90$ states/(eV·nm$^3$) \cite{private2025}, corresponding to a driving frequency of approximately 1.3 GHz---typical of elliptical SRF cavities. The simulation domain is a cube of side length $20\lambda$. In these calculations, we treat the normal conductivity $\sigma_n$ as a free parameter, and define the dimensionless ratio $\sigma_n / \sigma_{rt}$, where $\sigma_{rt} = 6.7 \times 10^6$ S/m is the room-temperature conductivity of Nb.

\begin{figure}
  \includegraphics[width=1.0\columnwidth]{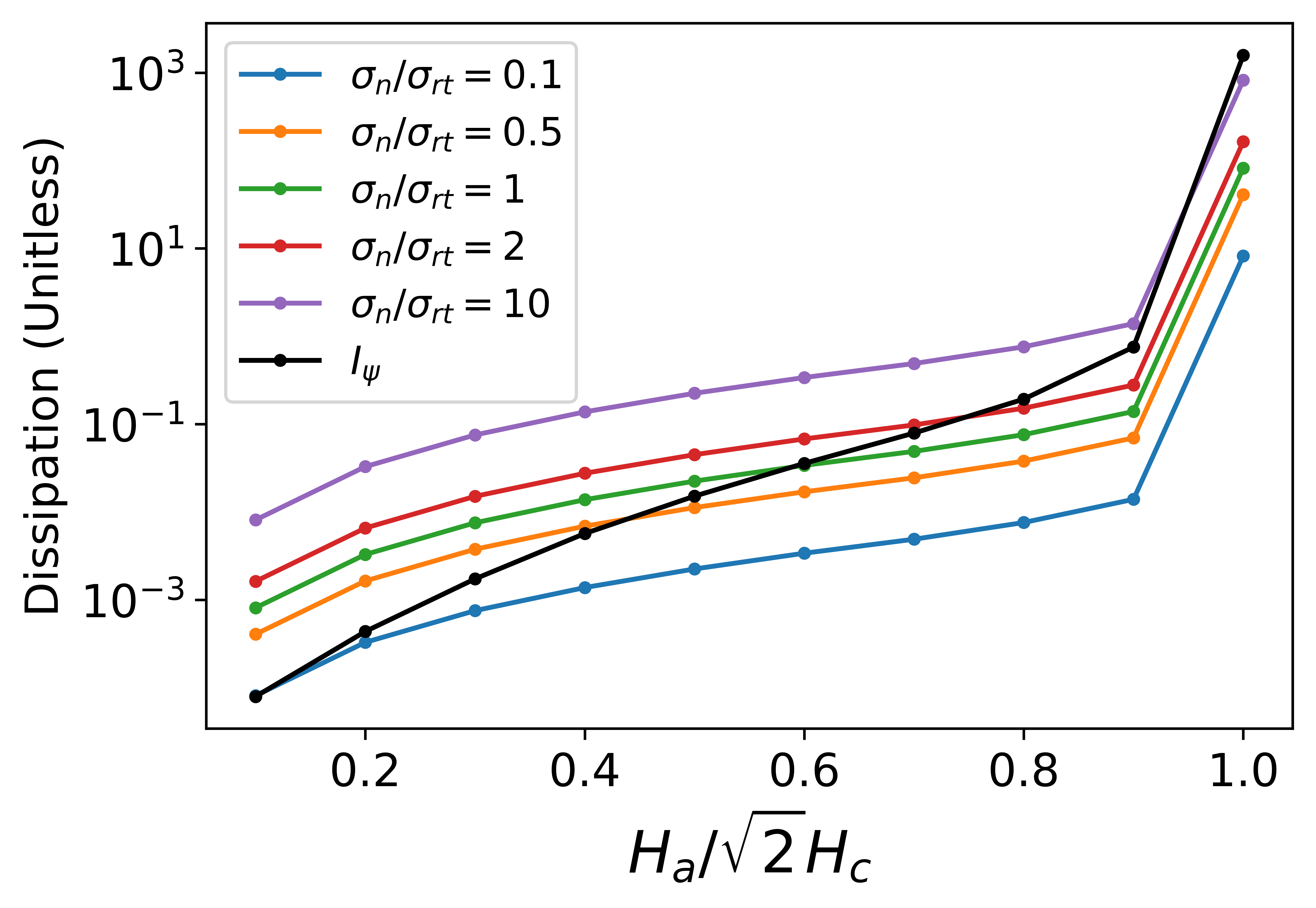}
  \caption{\justifying \label{fig:Nb_Dissipation_Comparison}
    \textbf{Dissipation Contributions from Current Response and Order Parameter Dynamics.}
    The two dissipation terms from Equation~\ref{surface_resistance_eq} are plotted separately as functions of applied field, for several values of the normal conductivity $\sigma_n$, shown here as ratios to the room-temperature conductivity $\sigma_{rt}$ of Nb. The term proportional to $\sigma_n$ and involving $I_A$ represents dissipation from normal current response, while the $I_\psi$ term reflects dissipation due to relaxation of the superconducting order parameter. As field increases, $I_\psi$ grows more rapidly and eventually overtakes $I_A$. The crossover point depends on the value of $\sigma_n / \sigma_{rt}$: for small values (e.g., $0.1$), $I_\psi$ dominates even at low field, while for large values (e.g., $10$), the current-induced dissipation remains dominant until vortex nucleation occurs near the superheating field $H_{sh}$. While $\sigma_n$ is varied here as a free parameter, this analysis helps illustrate how the relative importance of the two dissipation mechanisms depends on conductivity and field strength within the TDGL framework.
  }
\end{figure}

Figure~\ref{fig:Nb_Dissipation_Comparison} plots the two dissipation terms as functions of applied field for several values of $\sigma_n / \sigma_{rt}$. The $I_A$ term scales linearly with $\sigma_n$, while $I_\psi$ is independent of conductivity and increases more steeply with field. Although we treat $\sigma_n$ as a free parameter in these simulations, we are nominally modeling Nb, for which a physically meaningful value of $\sigma_n$ exists. Rigorously determining this value at cryogenic temperatures and GHz frequencies lies beyond the scope of this paper, but physical intuition provides useful bounds: at low field, the BCS-like (i.e., $I_A$) term should dominate in clean Nb cavities, while at high field, particularly once vortices nucleate, the residual-like (i.e., $I_\psi$) term should become dominant. These trends imply that the true conductivity is unlikely to differ from $\sigma_{rt}$ by more than an order of magnitude in either direction. This observation motivates our use of $\sigma_n = \sigma_{rt}$ in the simulations presented in the following section.

\begin{figure}
  \includegraphics[width=1.0\columnwidth]{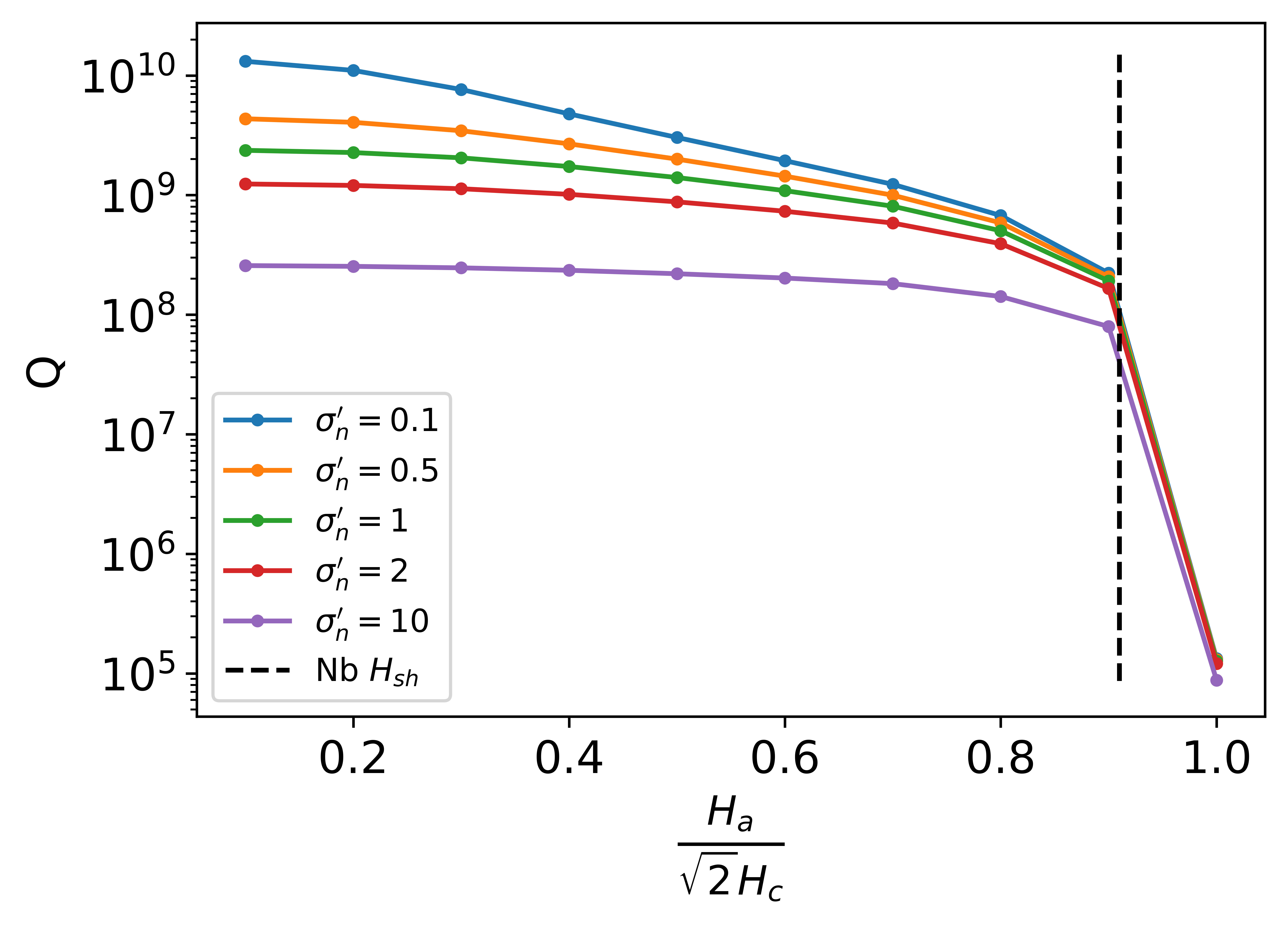}
  \caption{\justifying \label{fig:Nb_Q_vs_Hsh}
    \textbf{Nb SRF Cavity Quality Factor vs. Applied Field.}
    A plot of quality factor for a uniform Nb simulation versus the applied field (in units of $\sqrt{2}H_c$) for several different values of the normal conductivity, given by $\sigma_n =\sigma_{n1}\cdot \sigma_n'$, where $\sigma_{n1}$ is the room temperature conductivity of Nb. The quality factor is calculated using Eqs. \ref{reduced_quality_factor_eq} and \ref{surface_resistance_eq} with $G = 270$ $\Omega$. The calculation for several different values of $\sigma_n'$ change the low field value, and the shape of the curves, but has much less of an impact near $H_{sh}$. The dotted black line indicates the superheating field of Nb. Because it is above $H_{sh}$, vortex nucleation occurred for $\Tilde{H_a} = 1.0$ which is the reason for the sharp drop in $Q$.
  }
\end{figure}

While our primary emphasis is on the dissipation terms themselves, we can also use them to estimate quality factors via Eqs.~\ref{reduced_quality_factor_eq} and \ref{surface_resistance_eq}. As shown in Figure~\ref{fig:Nb_Q_vs_Hsh}, the shape of the resulting $Q$ curves depends on the choice of $\sigma_n / \sigma_{rt}$. For moderate and large values (e.g., 1 or 10), the quality factor remains relatively flat at low fields and drops sharply near the superheating field—consistent with expectations for clean Nb cavities. In contrast, for small conductivity values (e.g., 0.1), $Q$ decreases too rapidly with field, indicating unrealistic dissipation behavior. Although lower $\sigma_n$ values shift the low-field $Q$ magnitude closer to the experimental range of $10^{10}$–$10^{11}$, this comes at the cost of distorting the overall field dependence. Due to the gapless nature of TDGL, we generally expect $Q$ to be underestimated. While we do vary $\sigma_n$ in these calculations, it is not used as a fitting parameter in the traditional sense—namely, we do not select $\sigma_n$ to match the low-field $Q$ magnitude from experiment, as we see that doing so would yield qualitatively inaccurate results. Instead, we explore a range of values to identify physically plausible regimes and to qualitatively interpret the resulting dissipation behavior.

These reference simulations validate the dissipation decomposition and demonstrate how our framework reproduces key qualitative features of SRF performance. In the following section, we apply the same methodology to simulations of Sn-deficient islands in Nb\textsubscript{3}Sn to investigate how localized stoichiometric defects affect the balance between current-induced and order parameter–related dissipation. Building on this, we introduce an approach for estimating cavity-level losses by linking local vortex nucleation behavior to global quality factor trends.

\subsection{Dissipation and Quality Factor for Nb\textsubscript{3}Sn cavities with Sn-deficient Islands} \label{section_sn_islands_quality_factor}

To explore how localized stoichiometric defects alter RF dissipation, we now apply our framework to a representative simulation of a Sn-deficient island in Nb\textsubscript{3}Sn. This case illustrates how such inhomogeneities can shift the balance between current-induced and order parameter–related losses.

To evaluate dissipation in the presence of Sn-deficient islands, we simulate a time-dependent applied field with period $1000\tau_{\psi_0}$, corresponding to a frequency of approximately 5GHz. This is higher than the $\sim$1.3GHz typically used in Nb\textsubscript{3}Sn cavities, but was chosen to reduce simulation time for these computationally intensive cases. Each simulation required roughly five weeks of continuous runtime on a computing cluster, making longer periods prohibitively expensive. While quantitative estimates of dissipation will differ at lower frequency, we expect the qualitative features of the results to remain representative. To manage computational cost, we model a single island with radius $0.5\lambda$ located $0.5\lambda$ beneath the surface, using a reduced domain size of $5\lambda \times 2.5\lambda \times 2.5\lambda$ oriented to align the applied field and potential vortex motion along the $x$-axis.

\begin{figure}
    \includegraphics[width=1.0\columnwidth]{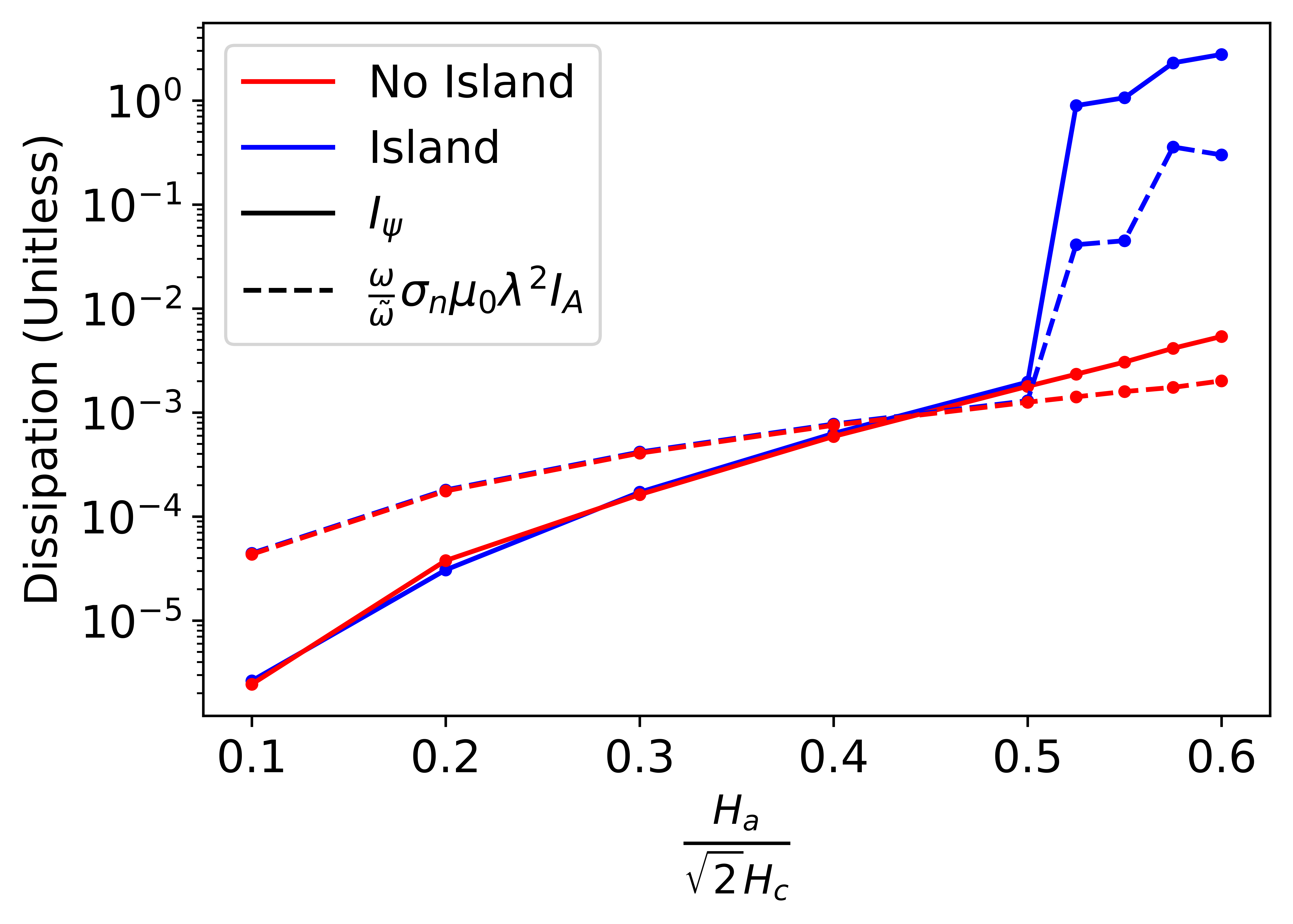}
    \caption{\justifying \label{fig:Sn_Islands_Q}
    \textbf{Dissipation Terms for Sn-deficient Island Simulations.}
    Dissipation terms $I_\psi$ (solid lines) and $I_A$ (dotted lines) from Eq.~\ref{surface_resistance_eq} are plotted as functions of applied field for simulations with (blue) and without (red) a Sn-deficient island. The ellipsoidal island has in-plane radii of $0.5\lambda$, a vertical radius of $0.25\lambda$, and its top edge located $0.5\lambda$ beneath the surface. In the island case, vortices nucleate at $H_a \approx 0.525$, causing both dissipation terms—especially $I_\psi$—to increase sharply by several orders of magnitude. In contrast, the no-island simulation exhibits smoother behavior with more modest growth, and $I_\psi$ exceeds $I_A$ by a smaller factor (typically $2$–$5\times$) at high field. These results highlight how local vortex penetration dramatically enhances both forms of dissipation, with $I_\psi$ becoming the dominant contribution.}
\end{figure}

Figure~\ref{fig:Sn_Islands_Q} shows the dissipation terms $I_\psi$ and $I_A$ for simulations with and without a Sn-deficient island. Below the vortex penetration field ($H_{vort} \approx 0.525$), both simulations yield nearly identical dissipation, indicating that embedded islands have minimal effect in the Meissner state. Once $H_a$ exceeds $H_{vort}$, however, the island simulation exhibits an abrupt and sustained increase in both dissipation terms—particularly in $I_\psi$, which becomes the dominant contribution and rises by several orders of magnitude. The dissipation values above $H_{vort}$ also appear somewhat noisy, which we attribute to numerical variations in vortex entry and dynamics. This binary behavior---where dissipation remains low until vortex entry---appears characteristic of embedded Sn-deficient regions in Nb\textsubscript{3}Sn. Whether other types of defects, such as surface-connected grain boundaries, show similar trends remains an open question. Exploring such cases is a clear next direction for future studies.

Building on this result, we develop an approximate method for estimating how a distribution of Sn-deficient islands could collectively affect the quality factor. The dissipation behavior observed in Figure~\ref{fig:Sn_Islands_Q} suggests that embedded islands have negligible impact below their respective vortex penetration fields, but trigger a sharp increase in dissipation once vortices begin to nucleate. This motivates a two-state approximation: for each island geometry, we assume dissipation follows the defect-free case up to $H_{vort}$, and then transitions abruptly to constant, elevated values beyond that point. For the vortex state, we use fixed dissipation values of $I_\psi = 2$ and $\frac{\omega}{\tilde{\omega}} \sigma_n \mu_0 \lambda^2 \cdot I_A = 0.1$, based on the average values from the simulation in Figure~\ref{fig:Sn_Islands_Q}.

\begin{figure}
  \includegraphics[width=1.0\columnwidth]{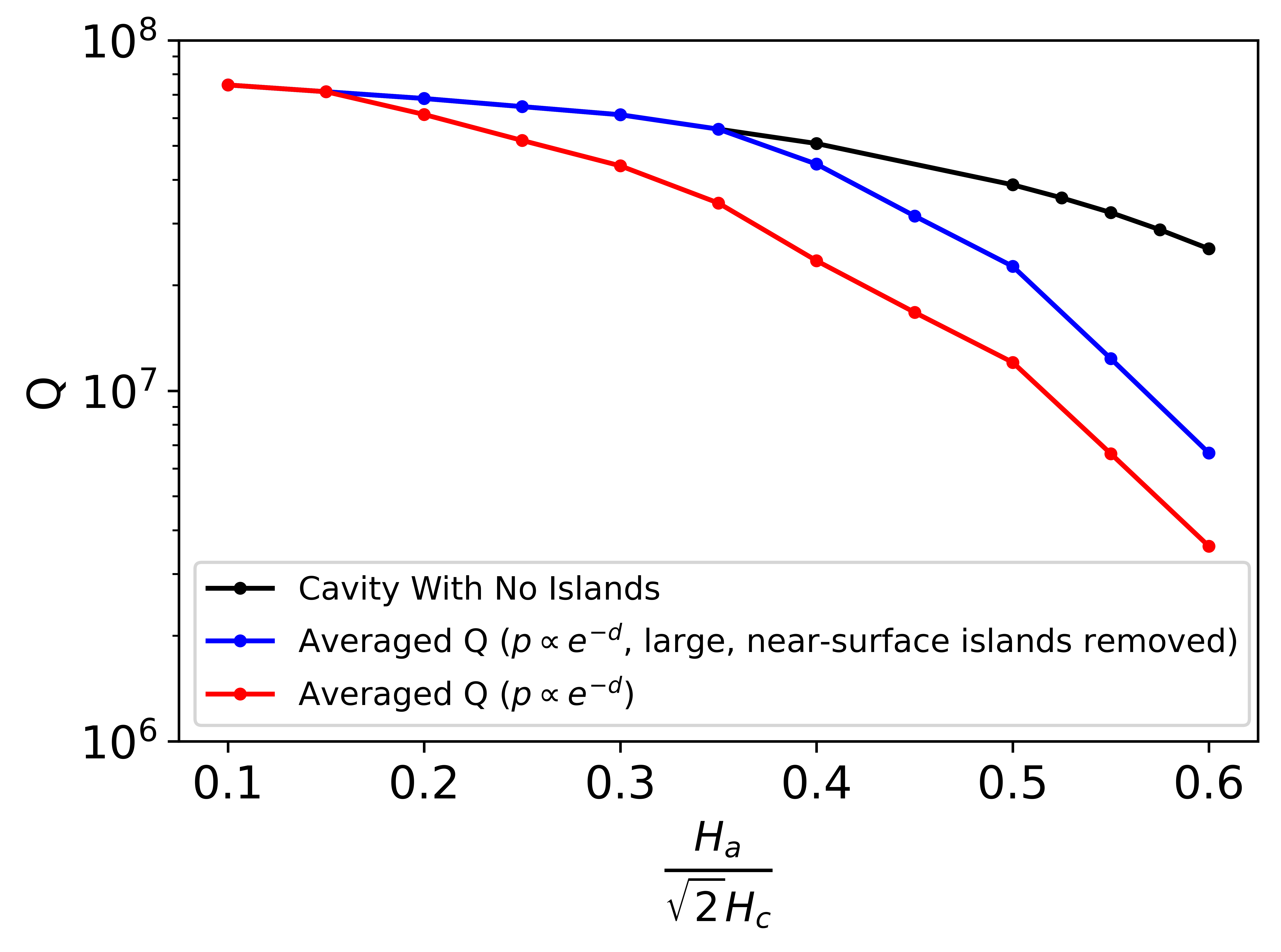}
  \caption{\justifying \label{fig:Hypothetical_Q_Slopes}
    \textbf{Estimated quality factor curves based on weighted dissipation from different Sn-deficient islands.}
    The black curve shows the quality factor for a defect-free cavity. The red curve represents a hypothetical quality factor obtained by averaging dissipation profiles from multiple Sn-deficient island geometries, each transitioning to fixed dissipation values of $I_\psi = 2$ and $\frac{\omega}{\tilde{\omega}} \sigma_n \mu_0 \lambda^2 \cdot I_A = 0.1$ above their respective $H_{vort}$. The weighting assumes all island sizes are equally likely and that the depth $d$ from the surface to the island edge follows an exponential distribution, $p(d) \propto e^{-d}$, normalized so the weights sum to one. The blue curve uses the same assumptions, except that $p(d) = 0$ for the $r = 0.5\lambda$ island when $d \leq 0.1\lambda$, and for the $r = \lambda$ and $r = 2\lambda$ islands when $d < \lambda$. All remaining weights are re-normalized so the distribution still sums to unity. Removing these large, near-surface defects in the model delays the onset of dissipation and improves high-field performance.}
\end{figure}

Using this approximation, we construct dissipation profiles for each of the simulated island geometries shown in Figure~\ref{fig:Sn_Islands_Hsh}, focusing on the cases with Sn concentration 18.7\%. For each defect, the dissipation follows the defect-free simulation up to its corresponding vortex penetration field $H_{vort}$, and then transitions to constant values of $I_\psi = 2$ and $\frac{\omega}{\tilde{\omega}} \sigma_n \mu_0 \lambda^2 \cdot I_A = 0.1$ once vortices nucleate. To estimate the aggregate dissipation and resulting quality factor, we assign weights to each defect based on an assumed distribution and compute a weighted average of the dissipation terms. For the red curve in Figure~\ref{fig:Hypothetical_Q_Slopes}, we assume all three island sizes are equally likely and that the depth $d$ from the surface to the outer edge of the island follows an exponential distribution $p(d) \propto e^{-d}$, normalized so the total probability sums to unity.

To evaluate the hypothetical benefit of eliminating large, near-surface defects, we also construct the blue curve in Figure~\ref{fig:Hypothetical_Q_Slopes} using a modified piecewise distribution. In this case, the probability $p(d)$ is defined as
\[
p(d) \propto 
\begin{cases}
0 & \text{for } r = 0.5\lambda \text{ and } d \leq 0.1\lambda \\
0 & \text{for } r = \lambda \text{ or } 2\lambda \text{ and } d < \lambda \\
e^{-d} & \text{otherwise}
\end{cases}
\]
and the resulting weights are renormalized to ensure they sum to one. Comparing the red and blue curves illustrates how removing such high-impact defects can delay the onset of increased dissipation and improve the high-field behavior of the cavity.

While the estimated quality factors in Figure~\ref{fig:Hypothetical_Q_Slopes} provide valuable qualitative insight, they should not be interpreted quantitatively. The low-field $Q$ magnitude in our simulations is just under $10^8$, significantly lower than the $\sim$$10^{10}$ typically observed in Nb$3$Sn cavities. This discrepancy is primarily due to the higher simulation frequency ($\sim$5 GHz versus 1.3 GHz in practice), though the quantitative limitations of TDGL---such as the assumption of gapless superconductivity and its reduced validity far below $T_c$---also contribute. Furthermore, we approximate the dissipation above $H_{vort}$ as constant, which neglects the increasing losses and possible thermal effects that would realistically occur as more vortices enter. In practice, such heating could trigger early cavity quench, while in our simulations the dissipation simply continues to increase with field without ever initiating a quench. Despite these simplifications, the structure of the estimated quality factors still captures important qualitative features.

In particular, the onset of high-field $Q$-slope in these plots arises directly from vortex nucleation, a steady-state property with broader validity in the dirty limit where TDGL is more appropriate. While the precise drop in $Q$ is uncertain, the onset of declining $Q$ in our aggregated curves is governed by the lowest vortex penetration field ($H_{vort}$) among the defect types with nonzero probability, highlighting how vortex entry from local defects could plausibly contribute to high-field $Q$-slope in at least some Nb$3$Sn cavities. More broadly, our results demonstrate a mechanism in which HFQS can emerge from the collective behavior of many small defects. Each defect follows a binary dissipation pattern---remaining nearly inert below $H_{vort}$ and sharply increasing in loss once vortex nucleation occurs---and as the field increases, more defects transition into this high-loss state. The aggregated effect produces a decline in $Q$ that qualitatively resembles the high-field $Q$-slope observed experimentally. While this is unlikely to be the sole mechanism underlying HFQS, it offers a plausible explanation for how distributed sub-surface defects may contribute to performance degradation in certain cavities, as we elaborate further in the conclusion.

\section{Conclusion}
\label{Conclusions}

In this paper, we have demonstrated a framework for incorporating sample-specific information into time-dependent Ginzburg–Landau (TDGL) simulations of superconducting radio-frequency (SRF) cavities. While TDGL is inherently suited for mesoscopic modeling, its application to non-uniform materials often involves arbitrary parameter choices. Our approach addresses this gap by linking TDGL parameters to well-defined microscopic material properties, allowing simulations to reflect characteristics obtained from density functional theory (DFT) or experimental measurements. This enables more realistic modeling of sample-specific features while preserving the computational tractability of TDGL on mesoscopic domains.

In addition to the sample-specific framework, we introduced a method for estimating dissipation and quality factor directly from TDGL simulations. While these estimates are subject to limitations---such as TDGL's quantitative inaccuracy far below $T_c$ and the computational necessity of simulating at higher frequencies---they offer a valuable qualitative connection between mesoscopic defect behavior and macroscopic performance degradation. By decomposing dissipation into physically meaningful components and linking vortex nucleation to high-field $Q$-slope, our approach provides insight into how distributed material inhomogeneities may impact SRF cavity performance. This type of dissipation-based analysis remains relatively uncommon in the TDGL literature and may serve as a useful tool for future studies aiming to bridge simulation and experiment.

We applied our framework to model Sn-deficient islands in Nb\textsubscript{3}Sn and found that they can reduce the vortex penetration field by as much as 60\% when located near the surface. This suggests that such sub-surface defects may play a role in limiting the achievable accelerating gradients in Nb\textsubscript{3}Sn cavities, particularly given that experimentally observed gradients remain well below the theoretical maximum. To explore this further, we computed the dissipation for representative island configurations and observed sharp increases in both dissipation terms above $H_{vort}$ due to vortex-induced losses. These effects are expected to be even more pronounced for larger or more exposed defects, and in future simulations that incorporate thermal feedback to model cavity quenching. Our method can also be readily extended to other defect types, such as grain boundaries, which are widely believed to impact cavity performance.

To illustrate how these mesoscopic effects might collectively influence macroscopic cavity behavior, we constructed hypothetical quality factor curves by assigning defect-specific $H_{vort}$ thresholds and aggregating their dissipation using simple assumptions about defect size and depth distributions. The resulting $Q$ curves qualitatively resemble the high-field $Q$-slope observed in experimental measurements, offering a plausible explanation in which HFQS emerges from the collective activation of many small, embedded defects. This analysis highlights the potential of TDGL-based simulations to inform defect mitigation strategies and guide the development of higher-performance SRF materials.

A recent experimental study by Viklund et al.~\cite{viklund2023_CBP} found that applying centrifugal barrel polishing (CBP) to Nb\textsubscript{3}Sn SRF cavities decreased the overall quality factor, lowered the field at which $Q$-slope began, and reduced the quench field. After an additional Sn vapor deposition, however, the cavity recovered its previous $Q$ performance and reached a higher quench field. Our results offer a potential explanation for these findings. CBP is known to smooth the cavity surface but also removes surface material, which may expose previously buried Sn-deficient islands. As shown in our simulations, bringing such defects closer to the surface significantly enhances their dissipation and lowers $H_{vort}$, potentially triggering early vortex entry, high-field $Q$-slope, and premature quenching. The follow-up Sn coating likely fills in these exposed islands, eliminating the source of additional dissipation. This healing process restores $Q$ and enables higher accelerating gradients, consistent with our finding that suppressing near-surface Sn-deficient regions improves both quality factor and vortex-related performance limits.

Whenever TDGL equations are used to model experimental systems, it is important to consider their limitations. The standard TDGL formalism is derived under the assumption of gapless superconductivity, since a gapped density of states introduces a singularity that precludes expansion in powers of the energy gap \cite{tinkham2004introduction}. Additionally, the equations are only quantitatively valid near the superconducting critical temperature. The former limitation can be addressed by using a generalized TDGL formulation developed by Kramer and Watts-Tobin \cite{Kramer_gTDGL}, which extends validity to gapped superconductors. However, several studies have observed conditions in SRF cavities that support the use of conventional TDGL. For example, Proslier et al.\cite{Proslier_Surface_Impurities} reported a broadened density of states at the surface of Nb cavities due to oxide layers, resulting in gapless surface superconductivity. Further work by Gurevich and Kubo\cite{Gurevich_SRF_DOS,Kubo_SRF_DOS} demonstrated that typical material compositions and SRF operating conditions often produce a broadened density of states and suppressed energy gap, reinforcing the relevance of TDGL for SRF applications. Nevertheless, such gapless behavior may come at a cost: broader density of states near the surface could contribute to increased dissipation and reduced quality factors.

The methods presented here for simulating realistic, sample-specific defects and estimating their impact on quality factor provide a new framework for linking microscopic material features to macroscopic SRF performance. While the underlying simulations are conceptually straightforward, their successful application depends on two critical inputs: accurate theoretical models of a material’s microscopic properties, and detailed experimental characterization of the defects present in real samples. For this reason, our method is best suited to research contexts where both inputs are available---either in well-studied materials systems or through interdisciplinary collaborations like those found in the NSF Center for Bright Beams, which helped motivate this work. SRF research is particularly well positioned for such integration. SRF cavities are inherently macroscopic devices whose performance depends sensitively on microscopic structure, yet direct experimental probing of such features is often infeasible. The approach we describe enables mesoscopic-scale simulations rooted in real material data, offering a valuable tool for designing and optimizing future generations of SRF cavities.

\section{Acknowledgement}
This work was supported by the US National Science Foundation under Award OIA-1549132, the Center for Bright Beams. We would like to thank Dr. Nathan Sitaraman for helpful conversations and estimates of the Fermi level density of states and Fermi velocities, and Dr. Michelle Kelley for helpful comments about early drafts of the methods section.

\appendix*
\section{TDGL Non-dimensionalization}

To nondimensionalize the TDGL equations, we start with Eqs. \ref{psi_eq_units} and \ref{j_eq_units}, and make the following coordinate transformations:
\begin{align*}
    \nabla &\longrightarrow \frac{1}{\lambda_0}\Tilde{\nabla} \\
    \frac{\partial}{\partial t} &\longrightarrow \frac{1}{\tau_{\psi_0}} \frac{\partial}{\partial \Tilde{t}} \\
    \mathbf{A}  &\longrightarrow \sqrt{2}H_{c0}\lambda_0\Tilde{\mathbf{A}} \\
    \psi &\longrightarrow \sqrt{\frac{|\alpha_0|}{\beta_0}}\Tilde{\psi}\\
    \phi &\longrightarrow \phi_0\Tilde{\phi}.
\end{align*}
If we substitute in Eqs. \ref{alpha_scale}, \ref{beta_scale}, \ref{gamma_scale}, and \ref{sigma_scale} for $\alpha$, $\beta$, $\Gamma$, and $\sigma_n$ respectively, we can then define the quantities:
\begin{align*}
    \lambda_0 &= \sqrt{\frac{m_s c^2 \beta_0}{4\pi e^2_s |\alpha_0|}} \\
    \xi_0 &= \sqrt{\frac{\hbar^2}{2m_s|\alpha_0|}} \\
    \kappa_0 &= \frac{\lambda_0}{\xi_0} \\
    H_{c0} &= \sqrt{\frac{4\pi\alpha^2_0}{\beta_0}} \\
    \tau_{\psi_0} &= \frac{\Gamma_0}{|\alpha_0|} \\
    \tau_{j_0} &= \frac{\sigma_{n0}m_s\beta_0}{e_s^2|\alpha_0|} \\
    u_0 &= \frac{\tau_{\psi_0}}{\tau_{j_0}} \\
    \phi_0 &= \frac{\hbar \kappa_0}{e_s \tau_{\psi_0}}.
\end{align*}
Using these relations, the resulting equations under the above coordinate transformations simplify into Eqs. \ref{psi_eq} and \ref{current_eq} (where we then drop the tildes).

\section{Approximate Estimation of Normal-State Conductivity}

We outline here a possible route to estimate $\sigma_n$ using the Drude model \cite{ashcroft-mermin}, where the electrical conductivity is given by
\begin{equation}
\sigma = \frac{n e^2 \tau}{m},\label{drude_conductivity}
\end{equation}
with $e$ and $m$ the electron charge and mass, $n$ the carrier density, and $\tau$ the mean free collision time. For the normal quasiparticle density at low temperatures, we adopt the approximation from Ref. \citenum{oseroff2022advancing},
\begin{equation}
n_n = 8ne^{-\Delta/k_bT}, \label{normal_carrier_denstiy_eq}
\end{equation}
which leads to an expression for the normal-state conductivity in the Meissner regime:
\begin{equation}
\sigma_n = \frac{n e^2 \tau}{m}\left(8e^{-\Delta/k_bT}\right).
\end{equation}
We emphasize that this expression is approximate, and moreover that the energy gap $\Delta$ is inferred from the TDGL order parameter $\psi$, which is known to overestimate $\Delta$ at low temperatures \cite{maki1963persistent}. Given these limitations, we regard this estimation primarily as a qualitative reference and have proceed with treating $\sigma_n$ as a free parameter in the main text.

\section{Quality Factor Derivation} \label{Appendix:Q_Derivation}
\begin{figure}
  \centerline{\includegraphics[width=1.0\columnwidth]{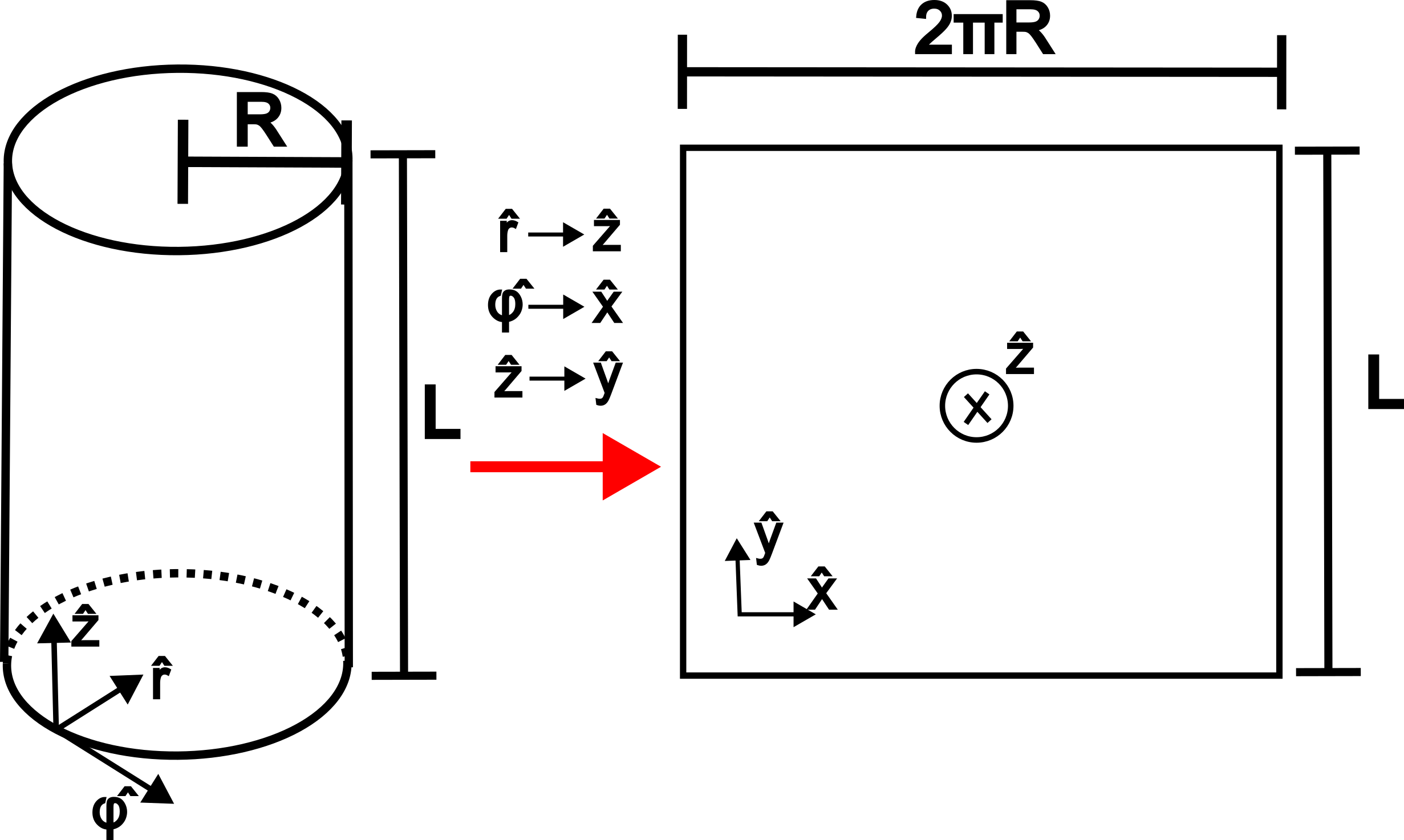}}
  \caption{\label{fig:Q_calculation_geometry}
    \textbf{Schematic of Transformations for the Quality Factor Calculation.}
     The cylindrical geometry is shown on the left, with the cavity radius $R$ and length $L$ depicted, and the coordinate directions, $\hat{r}$, $\hat{\phi}$, and $\hat{z}$. Under the transformation (on the right) the coordinates become cartesian, with the $\hat{r}$ direction becoming the new $\hat{z}$ direction, the $\hat{\phi}$ direction becomes the $\hat{x}$ direction, and the old $\hat{z}$ direction becomes the $\hat{y}$ direction.
  }
\end{figure}

We start with the quality factor:
\begin{equation}
    Q = \frac{2\pi E}{\Delta E}. \label{quality_factor_eq_appendix}
\end{equation}
These quantities (working in SI units for this section) can be expressed as integrals:
\begin{align}
    E = \frac{1}{2}\mu_0\int_V dV \mathbf{H}^2 \label{Energy_Integral} \\
    \Delta E = \int_0^T dt \int_{V_{surf}} dV_{surf} D, \label{Dissipated_Energy_Integral_appendix}
\end{align}
where $V$ is the cavity volume, $T$ is the RF period, and $D$ is given by Equation \ref{dissipation_eq}. $V_{surf}$ is the volume in the first few penetration depths of the cavity surface where essentially all of the dissipation occurs. TDGL simulation outputs are unit-free, so it is helpful to pull constants with units out of these integrals, leaving behind dimensionless functions which can be calculated from TDGL solutions. We start by expressing Equation \ref{Energy_Integral} in cylindrical coordinates:
\begin{equation*}
    E = \frac{1}{2}\mu_0 \int r dr \int d\phi \int dz \mathbf{H}^2
\end{equation*}
We then define some dimensionless quantities:
\begin{align}
    \Tilde{r} &= \frac{r}{R} \\
    \Tilde{z} &= \frac{z}{L} \\
    \Tilde{\mathbf{H}} &= \frac{\mathbf{H}}{H_a} 
\end{align}
Where $R$ is the maximum radius of the cavity, $L$ is the length of the cavity in the axial direction, and $H_a$ is the maximum value of the applied field at the surface of the cavity during an RF period.  These quantities allow the definition of a unit-less integral that only depends on the cavity geometry:
\begin{equation}
    I_H \equiv \int \Tilde{r}d\Tilde{r } \int d\Tilde{z} \Tilde{\mathbf{H^2}}
\end{equation}
Using these definitions with Equation \ref{Energy_Integral} and assuming that $\mathbf{H}$ has azimuthal symmetry results in
\begin{equation}
    E = \pi\mu_0 H_a^2 L R^2 I_H. \label{Cavity_Energy_Eq_appendix}
\end{equation}
Turning to the dissipated energy integral, suppose all of the dissipation occurs within a distance $d$ below the cavity surface, where $d << R$. This allows the cylindrical integral to be converted into cartesian coordinates, with the azimuthal direction becoming the new $x$ direction, the axial direction becoming the new $y$ direction, and the radial direction becoming the new $z$ direction. A diagram of these transformations is found in Figure \ref{fig:Q_calculation_geometry}.

With these transformations, we have
\begin{equation}
    \Delta E = \int_0^T dt \int_0^{2\pi R} dx \int_0^L dy \int_0^d dz D.
\end{equation}
When calculating this from simulation outputs, the integral is necessarily calculated over a small region of the overall cavity surface. Let $L_x$ and $L_y$ be the simulation domain size in the x and y directions respectively, and let $N$ be the total number of simulation areas needed to fully partition the cavity surface. Then the dissipation integral becomes
\begin{equation}
    \Delta E = N\int_0^T dt \int_0^{L_x} dx \int_0^{L_y} dy \int_0^d dz D,
\end{equation}
and $N$ can be approximated as
\begin{equation}
    N = \frac{2\pi R L}{L_x L_y}.
\end{equation}
Contuining as before we again define dimensionless coordinates:
\begin{align}
    \Tilde{x} &= \frac{x}{\lambda} \\
    \Tilde{y} &= \frac{y}{\lambda} \\
    \Tilde{z} &= \frac{z}{\lambda} \\
    \Tilde{t} &= \frac{T_{sim}}{T} t
\end{align}
where $\lambda$ is the penetration depth and $T_{sim}$ is the period in units of simulation time. These convert the integral to
\small\begin{equation}
    \Delta E = \lambda^3 \frac{T}{T_{sim}} N \int_0^{T_{sim}} d\Tilde{t} \int_0^{\frac{L_x}{\lambda}} d\Tilde{x} \int_0^{\frac{L_y}{\lambda}} d\Tilde{y} \int_0^{\frac{d}{\lambda}} d\Tilde{z} D.
\end{equation}\normalsize
Additionally, under the temporal gauge ($\phi = 0$), Equation \ref{dissipation_eq} can be expressed as
\small\begin{equation}
    D = 2\mu_0 H_c^2 \frac{T_{sim}}{T} \left(\left|\frac{\partial \Tilde{\psi}}{\partial \Tilde{t}}\right|^2 + \sigma_n \mu_0 \lambda^2 \frac{T_{sim}}{T} \left(\frac{\partial \Tilde{\mathbf{A}}}{\partial \Tilde{t}}\right)^2 \right), \label{dimension_separated_dissipation_appendix}
\end{equation}\normalsize
where $\Tilde{\psi}$ and $\Tilde{\mathbf{A}}$ are the unit-free versions of the vector potential and order parameter that are solved for with Eqs. \ref{psi_eq} and \ref{current_eq} (a derivation of Equation \ref{dimension_separated_dissipation_appendix} can be found in the next section of the Appendix). Finally, we define some more dimensionless integrals over the TDGL solutions:
\begin{align}
     I_\psi &\equiv \int_0^{T_{sim}} d\Tilde{t} \int_0^{\frac{L_x}{\lambda}} d\Tilde{x} \int_0^{\frac{L_y}{\lambda}} d\Tilde{y} \int_0^{\frac{d}{\lambda}} d\Tilde{z} \left|\frac{\partial\Tilde{\psi}}{\partial \Tilde{t}}\right|^2 \\
     I_A &\equiv \int_0^{T_{sim}} d\Tilde{t} \int_0^{\frac{L_x}{\lambda}} d\Tilde{x} \int_0^{\frac{L_y}{\lambda}} d\Tilde{y} \int_0^{\frac{d}{\lambda}} d\Tilde{z} \left(\frac{\partial\Tilde{\mathbf{A}}}{\partial \Tilde{t}}\right)^2
\end{align}
Combining everything and noting that $\omega = \frac{2\pi}{T}$, we get
\begin{equation}
    \Delta E = 2\mu_0 H_c^2\lambda^3 \frac{2\pi RL}{L_x L_y}\left(I_\psi + \omega \frac{\sigma_n \mu_0 \lambda^2 T_{sim}}{2\pi} I_A\right) \label{Dissipated_Energy_Eq_appendix}.
\end{equation}
Now using Equations \ref{quality_factor_eq_appendix}, \ref{Cavity_Energy_Eq_appendix}, and \ref{Dissipated_Energy_Eq_appendix} we get an expression for the quality factor,
\begin{equation}
    Q = \frac{\Tilde{H_a^2} R L_x L_y I_H}{2\lambda^3 \left(I_\psi + \omega \frac{\sigma_n \mu_0 \lambda^2 T_{sim}}{2\pi} I_A\right)},
\end{equation}
where $\Tilde{H_a} \equiv \frac{H_a}{\sqrt{2} H_c}$ is the applied field in simulation units. It is common to express the quality factor as
\begin{equation} \label{reduced_quality_factor_eq_appendix}
    Q = \frac{G}{R_s},
\end{equation}
where $R_s$ is the cavity surface resistance and $G$ is a geometric factor that depends only on quantities which are determined by the cavity geometry. We can define these quantities under the framework we have presented as
\begin{align}
    G &= \frac{1}{2}\mu_0 \omega R I_H\\
    R_s &= \frac{\mu_0 \omega \lambda^3}{\Tilde{H_a^2} L_x L_y}\left(I_\psi + \omega \frac{\sigma_n \mu_0 \lambda^2 T_{sim}}{2\pi} I_A\right). \label{surface_resistance_eq_appendix}
\end{align}
For a typical 1.3 GHz 9-cell Nb TESLA cavity, $G = 270$ $\Omega$ \cite{Nb_SRF_Cavity_Info}, so in practice we can just use this value or other known values of $G$, and only calculate $R_s$ from Equation \ref{surface_resistance_eq_appendix}.

\section{Nondimensionalizing the TDGL Dissipation}

We start with Equation \ref{dissipation_eq},
\begin{equation*}
    D = 2\Gamma\left|\left(\frac{\partial \psi}{\partial t} + \frac{ie_s\phi\psi}{\hbar}\right)\right|^2 + \sigma_n \mathbf{E}^2
\end{equation*}
choosing the temporal gauge ($\phi = 0$) we have
\begin{equation*}
    D = 2\Gamma\left|\frac{\partial \psi}{\partial t}\right|^2 + \sigma_n \left(\frac{\partial \mathbf{A}}{\partial t}\right)^2.
\end{equation*}
Next, we make the same coordinate transformations as from the previous section (and the same time transformation as from the methods section) and use the expressions for $\tau_\psi$ and $H_c$ on the first term:
\begin{align*}
    D &= \frac{2\Gamma \alpha}{\beta} \frac{T_{sim}^2}{T^2} \left|\frac{\partial \Tilde{\psi}}{\partial \Tilde{t}}\right|^2 + 2\sigma_n H_c^2 \lambda^2 \frac{T_{sim}^2}{T^2}\left(\frac{\partial \Tilde{\mathbf{A}}}{\partial \Tilde{t}}\right)^2 \\
    &= \frac{2 \tau_\psi \alpha^2}{\beta} \frac{T_{sim}^2}{T^2} \left|\frac{\partial \Tilde{\psi}}{\partial \Tilde{t}}\right|^2 + 2\sigma_n H_c^2 \lambda^2 \frac{T_{sim}^2}{T^2}\left(\frac{\partial \Tilde{\mathbf{A}}}{\partial \Tilde{t}}\right)^2 \\
    &= \frac{2 H_c^2}{4\pi} \frac{T_{sim}}{T} \left|\frac{\partial \Tilde{\psi}}{\partial \Tilde{t}}\right|^2 + 2\sigma_n H_c^2 \lambda^2 \frac{T_{sim}^2}{T^2}\left(\frac{\partial \Tilde{\mathbf{A}}}{\partial \Tilde{t}}\right)^2,
\end{align*}
where in the last line we used the fact that $T = \tau_\psi T_{sim}$. Finally, this expression is in Gaussian units so we convert to SI units so that it is compatible with the other expressions in Section \ref{Estimating_Quality_Factor}:
\begin{align*}
    D &= 2 \tau_\psi \mu_0 H_c^2 \frac{T_{sim}^2}{T^2} \left|\frac{\partial \Tilde{\psi}}{\partial \Tilde{t}}\right|^2 + 2\sigma_n \mu_0^2 H_c^2 \lambda^2 \frac{T_{sim}^2}{T^2}\left(\frac{\partial \Tilde{\mathbf{A}}}{\partial \Tilde{t}}\right)^2 \\
    &= 2\mu_0 H_c^2 \frac{T_{sim}}{T} \left(\left|\frac{\partial \Tilde{\psi}}{\partial \Tilde{t}}\right|^2 + \sigma_n \mu_0 \lambda^2 \frac{T_{sim}}{T} \left(\frac{\partial \Tilde{\mathbf{A}}}{\partial \Tilde{t}}\right)^2 \right)
\end{align*}

\bibliography{main_revised_2}

\end{document}